\documentclass[aps,pre,twocolumn,showpacs]{revtex4}
\usepackage{amsfonts}
\usepackage{amsmath}
\usepackage{amssymb}
\usepackage{graphicx}

\def\kbar{\protect\@kbar}
\def\@kbar{\relax \bgroup
\def\@tempa{\hbox{\raise.73\ht0
\hbox to0pt{\kern.25\wd0\vrule width.5\wd0 height.1pt
depth.1pt\hss}\box0}}\mathchoice{\setbox0\hbox{$\displaystyle
k$}\@tempa}{\setbox0\hbox{$\textstyle
k$}\@tempa}{\setbox0\hbox{$\scriptstyle
k$}\@tempa}{\setbox0\hbox{$\scriptscriptstyle k$}\@tempa}\egroup}

\begin{document}

\title{\textbf{Generic Superweak Chaos Induced by Hall Effect}}
\author{Moti Ben-Harush and Itzhack Dana}
\affiliation{Minerva Center and Department of Physics, Bar-Ilan University, Ramat-Gan
52900, Israel}

\begin{abstract}
We introduce and study the ``kicked Hall system" (KHS), i.e., charged
particles periodically kicked in the presence of uniform magnetic ($\mathbf{B%
}$) and electric ($\mathbf{E}$) fields that are perpendicular to each other
and to the kicking direction. We show that for resonant values of $B$ and $E$
and in the weak-chaos regime of sufficiently small nonintegrability
parameter $\kappa$ (the kicking strength), there exists a \emph{generic}
family of periodic kicking potentials for which the Hall effect from $%
\mathbf{B}$ and $\mathbf{E}$ significantly suppresses the weak chaos,
replacing it by \emph{``superweak"} chaos (SWC). This means that the system
behaves as if the kicking strength were $\kappa ^2$ rather than $\kappa$.
For $E=0$, SWC is known to be a classical fingerprint of quantum
antiresonance but it occurs under much less generic conditions, in
particular only for very special kicking potentials. Manifestations of SWC
are a decrease in the instability of periodic orbits and a narrowing of the
chaotic layers, relative to the ordinary weak-chaos case. Also, for global
SWC, taking place on an infinite ``stochastic web" in phase space, the
chaotic diffusion on the web is much slower than the weak-chaos one. Thus,
the Hall effect can be relatively stabilizing for small $\kappa$. In some
special cases, the effect is shown to cause ballistic motion for almost all
parameter values. The generic global SWC on stochastic webs in the KHS
appears to be the two-dimensional closest analog to the Arnol'd web in
higher dimensional systems.
\end{abstract}

\maketitle

\begin{center}
\textbf{I. INTRODUCTION}
\end{center}

The nature of chaotic transport in typical Hamiltonian systems is known to
depend on the system dimensionality \cite{chaos,bc}. For the
lowest-dimensional systems which can be nonintegrable, i.e., either
one-dimensional time-dependent systems or two-dimensional time-independent
ones (both described by area-preserving Poincar\'{e} maps),
Kolmogorov-Arnol'd-Moser (KAM) tori \cite{k,a,m} are barriers to chaotic
transport and thus form boundaries of localized chaotic regions. Only when
all the KAM tori break for sufficiently large nonintegrability parameter,
these regions merge into a global chaotic region permeating all the phase
space. A paradigmatic realistic model of this scenario is the famous kicked
rotor described by the Taylor-Chirikov standard map \cite{bc,jmg,mmp,df,dr}.
The situation is fundamentally different in higher-dimensional systems.
Because of purely topological reasons, KAM tori in these systems do not
divide phase space. Then, global chaotic transport on the so-called
``Arnol'd web" takes place generically for arbitrarily small
nonintegrability parameter but at a very slow rate \cite{a1,akn,kl}.

There is an analog to the Arnol'd web in area-preserving maps describing
another well-known paradigmatic system basically different from the kicked
rotor. This is the system of charged particles periodically kicked
perpendicularly to a uniform magnetic field \cite{wm,lw,y,da,dh,d,prk}.
Assuming, by proper choice of units and without loss of generality,
particles of unit mass and charge, the system is defined by the general
Hamiltonian: 
\begin{equation}  \label{H0}
H_0=\frac{\mathbf{\Pi}}{2}^2+KV(x) \sum_{s=-\infty}^{\infty}\delta(t-sT ), \
\ \ 
\end{equation}
where $\mathbf{\Pi}= \mathbf{p}-\mathbf{B}\times\mathbf{r}/(2c)$ is the
kinetic momentum in a uniform magnetic field ${\mathbf{B}}$ along the $z$
axis, $K$ is a nonintegrability parameter, $V(x)$ is a general periodic
potential, and $T$ is the time period. Let us summarize the relevant
properties of the system (\ref{H0}), see more details in Secs. II and IIIB.
This system is equivalent to a periodically kicked harmonic oscillator on
the phase plane $(u=\Pi_x/\omega,v=\Pi_y/\omega )$, where $\omega=B/c$ is
the cyclotron angular velocity; in terms of $v$, the potential reads $%
V(x)=V(x_{\mathrm{c}}-v)$ \cite{da}, where $x_{\mathrm{c}}$ is the $x$
coordinate of the cyclotron orbit center and is a constant of the motion.
KAM theory is not applicable to this system for small $K$ since the
harmonic-oscillator Hamiltonian is degenerate, being linear in the action.
In fact, at least for some values of $\omega T$, the Poincar\'{e} map for
the system has no KAM tori and exhibits a global weak chaos on a
``stochastic web" over all the phase space for arbitrarily small $\kappa
=K/\omega$ \cite{wm,da}, see Fig. 1. This web is thus analogous to the
Arnol'd web. For special values of $x_{\mathrm{c}}$ such that $V(x_{\mathrm{c%
}}-v)$ is an \emph{odd} function of $v$ (up to an additive constant), as in
Fig. 1(b), the width of the stochastic web and the diffusion rate on it are
much smaller than those for $V(x_{\mathrm{c}}-v)$ with generic values of $x_{%
\mathrm{c}}$ \cite{d,prk}; compare the diffusion rates in Figs. 1(a) and
1(b). These rare phenomena, manifestations of what we call \emph{``superweak
chaos"} (SWC), where originally discovered in Ref. \cite{d} as classical
fingerprints of quantum antiresonance; see also work \cite{dd}. SWC is due
to the fact that for small $\kappa$ the system behaves as if the
nonintegrability parameter were $\kappa^2$ rather than $\kappa$ \cite{d,prk}%
. 
\begin{figure}[tbp]
\centering
\includegraphics[width=9cm,natwidth=850,natheight=400]{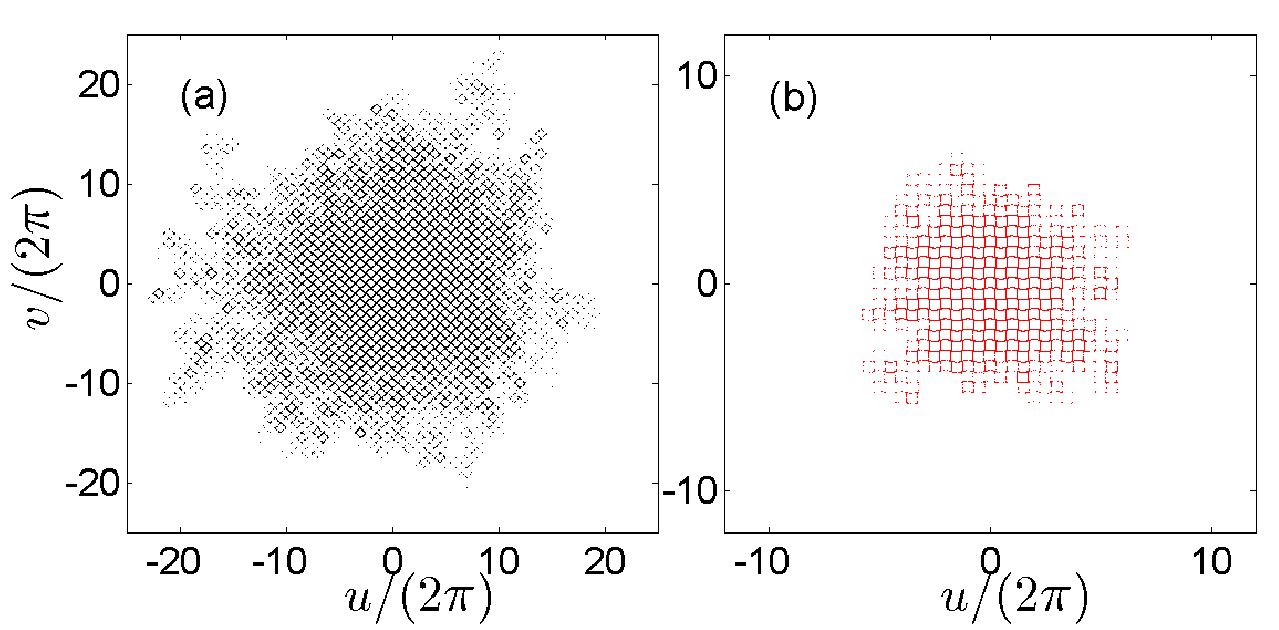}
\caption{(Color online) Global stochastic-web diffusion of a $20\times 20$
chaotic ensemble after 120000 iterations of the one-period map for the
system (\protect\ref{H0}) [i.e., the map (\protect\ref{Mh}) for $\protect%
\eta =0$] with $\protect\kappa =K/\protect\omega =0.6$, $V(x)=-\cos (x)$ [$%
f(x)=-\sin(x)$], $\protect\gamma =\protect\omega T=\protect\pi /2$, and two
values of the constant of the motion $x_{\mathrm{c}}=x_{\mathrm{c}}^{(0)}$:
(a) $x_{\mathrm{c}}=0$ [the effective potential $V(x_{\mathrm{c}}-v)=-\cos
(v)$ is even]; (b) $x_{\mathrm{c}}=\protect\pi /2$ [$V(x_{\mathrm{c}%
}-v)=-\sin (v)$ is odd]. Clearly, the diffusion in case (b) is significantly
slower than that in case (a).}
\label{fig1}
\end{figure}

In this paper, we introduce and study the ``kicked Hall system" (KHS), i.e.,
the system (\ref{H0}) with the addition of a uniform electric field ${%
\mathbf{E}}$ perpendicular to both ${\mathbf{B}}$ and the kicking direction.
We show that for resonant values of $B$ and $E$, defined by Eqs. (\ref{rge})
below, and for sufficiently small $\kappa$ the Hall effect from ${\mathbf{B}}
$ and ${\mathbf{E}}$ causes SWC, either local or global, to occur for a 
\emph{generic} family of kicking potentials. This is in contrast with the $%
E=0$ case, where SWC occurs under much more restrictive conditions, in
particular only for odd potentials in this family. Thus, the Hall effect
significantly stabilizes the system by transforming the weak chaos for $E=0$
into SWC. When the KHS has translational invariance in phase space, global
SWC on stochastic webs is shown to emerge under generic conditions while
ballistic motion occurs only in some special cases. The generic global SWC
on stochastic webs in the KHS, with a much smaller transport rate than that
of ordinary weak chaos, appears to be the two-dimensional closest analog to
the Arnol'd web in higher dimensional systems. However, while the rate of
Arnol'd diffusion decreases exponentially with the nonintegrability
parameter \cite{a1,akn,kl}, the SWC-diffusion rate is expected to decrease
only algebraically with this parameter, as in the case of ordinary weak
chaos on stochastic webs \cite{lw}.

The paper is organized as follows. In Sec. II, we present the general KHS in
natural coordinates and derive its basic Poincar\'{e} map under resonance
conditions on $B$ and $E$. In Sec. III, we define SWC and derive general
conditions for it in both the $E=0$ case (Sec. IIIB) and the $E\neq 0$ case
(Sec. IIIC); these conditions clearly imply that SWC occurs generically for
resonant $E\neq 0$. In Sec. IIID, we show that SWC leads to a decrease of
the linear instability of periodic orbits and to a narrowing of the
corresponding chaotic layers, relative to the weak-chaos case. In Sec. IV,
we consider the KHS with translational invariance in phase space, leading to
a global SWC on stochastic webs. We provide numerical evidence for the
suppression of the chaotic-diffusion rate for $E\neq 0$ relative to the
weak-chaos one for $E=0$. Integrable effective Hamiltonians, giving the
skeleton of the stochastic webs, are derived. In Sec. V, we show that, in
special cases of weak chaos and SWC, ballistic motion occurs for almost all
parameter values while stochastic webs emerge only in small parameter
intervals. A summary and conclusions are presented in Sec. VI. Several
technical details appear in the Appendices.

\begin{center}
\textbf{II. THE KICKED HALL SYSTEM (KHS) AND ITS POINCAR\'{E} MAP}

\textbf{A. KHS in natural coordinates}
\end{center}

The KHS is defined by adding to (\ref{H0}) a uniform electric field ${%
\mathbf{E}}$ in the $y$ direction, i.e., perpendicularly to both ${\mathbf{B}%
}$ and $x$: 
\begin{equation}
H=\frac{\mathbf{\Pi }}{2}^{2}-Ey+KV(x)\sum_{s=-\infty }^{\infty }\delta
(t-sT),\ \ \   \label{H}
\end{equation}%
where we recall that unit mass and charge are assumed, without loss of
generality. Let us express (\ref{H}) in the two natural degrees of freedom
in a magnetic field \cite{jl}. These are the independent conjugate pairs $%
(x_{\mathrm{c}},y_{\mathrm{c}})$ (coordinates of the cyclotron-orbit center)
and $(u=\Pi _{x}/\omega ,v=\Pi _{y}/\omega )$, with $\omega =B/c$; here $u$
and $-v$ are, respectively, the $y$ and $x$ coordinates of the radius vector
of a cyclotron orbit, so that $x=x_{\mathrm{c}}-v$ and $y=y_{\mathrm{c}}+u$.
Using these relations and defining the variable $u^{\prime }=u-E/\omega ^{2}$%
, which we re-denote by $u$, the Hamiltonian (\ref{H}) can be expressed as
follows: 
\begin{equation}
H=\omega ^{2}(u^{2}+v^{2})/2-Ey_{\mathrm{c}}+KV(x_{\mathrm{c}%
}-v)\sum_{s=-\infty }^{\infty }\delta (t-sT),  \label{eKHO}
\end{equation}%
where an insignificant constant $E^{2}/(2\omega ^{2})$ was omitted. As one
can easily check, the conjugate pairs above have Poisson brackets $\{y_{%
\mathrm{c}},x_{\mathrm{c}}\}=\{u,v\}=1/\omega $, so that the Hamilton
equation for $x_{\mathrm{c}}$ is $\dot{x}_{\mathrm{c}}=-1/(\omega )\partial
H/\partial y_{\mathrm{c}}=E/\omega $. Thus, $x_{\mathrm{c}}$ evolves
linearly in time (Hall effect): 
\begin{equation}
x_{\mathrm{c}}=x_{\mathrm{c}}^{(0)}+J_{\mathrm{Hall}}t,\ \ J_{\mathrm{Hall}}=%
\frac{E}{\omega },  \label{xct}
\end{equation}%
where $J_{\mathrm{Hall}}$ is the Hall velocity. Inserting Eq. (\ref{xct}) in
Eq. (\ref{eKHO}), we see that the Hamiltonian (\ref{eKHO}) for the conjugate
pair $(u,v)$ is essentially that of a kicked harmonic oscillator with a time
modulated kicking potential $V(x_{\mathrm{c}}^{(0)}+J_{\mathrm{Hall}}t-v)$.
For $E=0$, $J_{\mathrm{Hall}}=0$ and $x_{\mathrm{c}}$ is a constant of the
motion.

\begin{center}
\textbf{B. The Poincar\'{e} map and its iterates}
\end{center}

From $\{ u,v\} =1/\omega$, the Hamilton equations for $(u,v)$ are $\dot{u}%
=(1/\omega)\partial H/\partial v$ and $\dot{v}=-(1/\omega)\partial
H/\partial u$, where $H$ is given by (\ref{eKHO}) with (\ref{xct}).
Integrating the latter equations from time $t=sT-0$ to time $t=(s+1)T-0$ ($s$
integer) and denoting $u_s=u(t=sT-0)$, $v_s=v(t=sT-0)$, we easily obtain the
one-period Poincar\'{e} map for the system: 
\begin{equation}  \label{Mh}
\begin{array}{ccc}
u_{s+1} & = & [u_s+\kappa f(x_{\mathrm{c}}^{(0)}+s\eta-v_s)]\cos(\gamma)+v_s%
\sin(\gamma), \\ 
v_{s+1} & = & -[u_s+\kappa f(x_{\mathrm{c}}^{(0)}+s\eta-v_s)]\sin(%
\gamma)+v_s\cos(\gamma),%
\end{array}%
\end{equation}
where $\kappa =K/\omega$, $\gamma=\omega T$, $\eta=J_{\mathrm{Hall}%
}T=ET/\omega $, and $f(x)=-dV/dx$ is the force function. For $E=0$, with
arbitrary constant value of $x_{\mathrm{c}}=x_{\mathrm{c}}^{(0)}$, the map (%
\ref{Mh}) reduces to a generalized version of the Zaslavsky web map \cite%
{wm,da}.

Defining $z_s=u_s+iv_s$, the map (\ref{Mh}) can be written more compactly
as: 
\begin{equation}  \label{cMh}
M_{\gamma ,\eta}:\ z_{s+1}=[z_s+\kappa f(x_{\mathrm{c}}^{(0)}+s%
\eta-v_s)]e^{-i\gamma}.
\end{equation}
After $s$ iterations of (\ref{cMh}) starting from $z=z_0$, we get: 
\begin{equation}  \label{cMhs}
M_{\gamma ,\eta ,s}:\ z_s=\left [ z_0+\kappa \sum_{j=0}^{s-1}f(x_{\mathrm{c}%
}^{(0)}+j\eta-v_j)e^{ij\gamma}\right ]e^{-is\gamma}.
\end{equation}

We choose length units such that the period of $V(x)$ is $2\pi$ and assume
from now on rational values of $\gamma /(2\pi )$ and $\eta /(2\pi )$: 
\begin{equation}  \label{rge}
\frac{\gamma}{2\pi}=\frac{m}{n},\ \ \ \ \frac{\eta}{2\pi}=\frac{k}{l},
\end{equation}
where $(m,n)$ and $(k,l)$ are two pairs of coprime integers. Due to the $2\pi
$-periodicity of both $\exp (-i\gamma )$ and $f(x)$ in Eq. (\ref{cMh}), Eqs.
(\ref{rge}) are resonant conditions on $\gamma$ and $\eta$: After the
minimal number $s=r$ of iterations, where $r=\mathrm{lcm}(n,l)$ is the least
common multiple of $n$ and $l$, one has $r\gamma \ \mathrm{mod}(2\pi )=r\eta
\ \mathrm{mod}(2\pi )=0$ and all the multiples of $\gamma$ and $\eta$ modulo 
$2\pi$ will appear in the map (\ref{cMhs}). In this sense, the map $%
M_{\gamma ,\eta ,s}$ ``closes" after not less than $s=r$ iterations, so that 
$M_{\gamma ,\eta ,r}$ may be considered as the basic map of the system under
the conditions (\ref{rge}): 
\begin{equation}  \label{cMhb}
M_{\gamma ,\eta ,r}:\ z_r=z_0+\kappa\sum_{j=0}^{r-1}f(x_{\mathrm{c}%
}^{(0)}+j\eta-v_j)e^{ij\gamma},\ \ r=\mathrm{lcm}(n,l).
\end{equation}
The importance of the map (\ref{cMhb}) is, in particular, that points of a
periodic orbit of (\ref{Mh}) are generally fixed points of some iterate of $%
M_{\gamma ,\eta ,r}$.

\begin{center}
\textbf{III. SUPERWEAK CHAOS (SWC)}

\textbf{A. General}
\end{center}

One can easily see that the map (\ref{Mh}) for $n=1,2$ ($\gamma =0,\pi$) is
integrable even for irrational $\eta /(2\pi )$. Thus, chaos can emerge only
for $n>2$. We then say that the map (\ref{cMhb}) for $n>2$ and small $%
\kappa\ll 1$ exhibits SWC if its expansion in powers of $\kappa$ starts from 
$\kappa ^2$, 
\begin{equation}  \label{SWC}
M_{\gamma ,\eta ,r}:\ z_r=z_0+O(\kappa ^2).
\end{equation}
This is unlike ordinary weak chaos, with $z_r=z_0+O(\kappa )$.

In Secs. IIIB and IIIC, we shall examine the validity of Eq. (\ref{SWC}) for
the general family of $2\pi$-periodic potentials with finite Fourier
expansion, 
\begin{equation}  \label{V}
V(x)=\sum_{g=-N}^NV_g\exp{(igx)},\ \ V_0=0,
\end{equation}
for both $E=0$ and $E\neq 0$. In Sec. IIID, we show that SWC leads to a
decrease of the linear instability of periodic orbits of the map (\ref{Mh})
and to a narrowing of the corresponding chaotic layers, relative to the
weak-chaos case.

\begin{center}
\textbf{B. SWC for }$\boldsymbol{E=0}$
\end{center}

We consider here the case of $E=0$ or $\eta =0$ in a framework more general
than in previous works \cite{d,prk}. For $\eta =0$, $x_{\mathrm{c}}=x_{%
\mathrm{c}}^{(0)}$ (constant) and $k/l=0/1$ in Eq. (\ref{rge}), so that $r=n$
in Eq. (\ref{cMhb}). We then show that Eq. (\ref{SWC}) holds only if $n$ is 
\emph{even} and the potential $V(x_{\mathrm{c}}-v)$ [with $V(x)$ given by
Eq. (\ref{V})] is \emph{odd}: $V(x_{\mathrm{c}}+v)=-V(x_{\mathrm{c}}-v)$.

Let us calculate the first-order term in the expansion of the map (\ref{cMhb}%
) in powers of $\kappa$ and determine under which conditions this term
vanishes. To this end, it is sufficient to calculate $v_j$ in Eq. (\ref{cMhb}%
) to zero order in $\kappa$. From Eq. (\ref{cMhs}) one has, to this order, $%
z_s=z_0\exp (-is\gamma )$. Using the latter relation together with Eq. (\ref%
{V}) and $f(x)=-dV/dx$, we get Eq. (\ref{cMhb}) (with $r=n$) to first order
in $\kappa$: 
\begin{equation}  \label{cMhb0}
z_n=z_0-i\kappa\sum_{g=-N}^N gV_g\exp (igx_{\mathrm{c}})S_{n,g}(u_0,v_0) ,
\end{equation}
\begin{eqnarray}
S_{n,g}(u_0,v_0) &=& \sum_{j=0}^{n-1}e^{2\pi i jm/n}\exp \{ig[u_0\sin (2\pi
jm/n)  \notag \\
&-& v_0\cos (2\pi jm/n)]\} .  \label{Sdn}
\end{eqnarray}
Equation (\ref{SWC}) will hold provided the coefficient of $\kappa$ in Eq. (%
\ref{cMhb0}) vanishes for all $(u_0,v_0)$. This is the case only if $%
gV_{g}\exp (igx_{\mathrm{c}})S_{n,g}(u_0,v_0)$ is an odd function of $g$ for
all $(u_0,v_0)$. This means that $V_{g}\exp (igx_{\mathrm{c}})$ and $%
S_{n,g}(u_0,v_0)$ are either both even or both odd functions of $g$. From
Eq. (\ref{Sdn}), we see that $S_{n,g}(u_0,v_0)$ has a definite parity under $%
g\rightarrow -g$ for all $(u_0,v_0)$ only if $n$ is even and then 
\begin{eqnarray}
S_{n,g}(u_0,v_0) &=& 2i\sum_{j=0}^{n/2-1}e^{2\pi i jm/n}\sin [gu_0\sin (2\pi
jm/n)  \notag \\
&-& gv_0\cos (2\pi jm/n)] ,  \notag
\end{eqnarray}
an odd function of $g$. Thus, $V_{g}\exp (igx_{\mathrm{c}})$ must be also
odd in $g$, implying $V(x_{\mathrm{c}}+v)=-V(x_{\mathrm{c}}-v)$.

\begin{center}
\textbf{C. Generic SWC for }$\boldsymbol{E\neq 0}$
\end{center}

Given arbitrary rational values of $\gamma/(2\pi )=m/n$ and $\eta /(2\pi
)=k/l\neq 0$ ($E\neq 0$) in Eq. (\ref{rge}), let us write $%
n/l=n^{\prime}/l^{\prime}$, where $(n^{\prime},l^{\prime})$ are coprime
integers. Consider the family of potentials (\ref{V}) with any given number $%
2N$ of harmonics. We then show here that if 
\begin{equation}  \label{cE}
l^{\prime}> N,
\end{equation}
SWC occurs for \emph{arbitrary} potential (\ref{V}) and initial value $x_{%
\mathrm{c}}^{(0)}$ in Eq. (\ref{cMhb}), irrespectively of the parity of $n$.
This is in contrast with the $E=0$ case for which SWC can occur only if $n$
is even and only if the potential (\ref{V}) and the constant $x_{\mathrm{c}}$
are such that $V(x_{\mathrm{c}}+v)$ is an odd function of $v$ (see Sec.
IIIB).

To show this, we first note that since $r=\mathrm{lcm}(n,l)$ one has $%
r=l^{\prime}n$. Let us write in Eq. (\ref{cMhb}) $j=bn+w$, where $b=0,\dots
,l^{\prime}-1$ and $w =0,\dots ,n-1$. Then, to zero order in $\kappa$, $%
z_j=z_0\exp (-ij\gamma)=z_{w}$ since $\gamma =2\pi m/n$; thus, $v_j=v_{w}$
to this order. Using all this in Eq. (\ref{cMhb}), we obtain, to first order
in $\kappa$: 
\begin{eqnarray}
z_r &=& z_0-i\kappa\sum_{w=0}^{n-1}e^{iw\gamma}\sum_{g=-N}^N gV_g\exp [ig(x_{%
\mathrm{c}}^{(0)}+w\eta-v_{w})]  \notag \\
&\times& \sum_{b=0}^{l^{\prime}-1}\exp (igbn\eta ) .  \label{cMhbe}
\end{eqnarray}
The second line in Eq. (\ref{cMhbe}) is a geometric series which vanishes
since $gl^{\prime}n\eta =2\pi gn^{\prime}k$ and $gn\eta =2\pi
gn^{\prime}k/l^{\prime}$ is never an integer multiple of $2\pi$ because of
condition (\ref{cE}). Thus, the coefficient of $\kappa$ in Eq. (\ref{cMhbe})
is identically zero, implying the SWC condition (\ref{SWC}).

As an example, Fig. 2 shows cases of $\gamma =2\pi /3$, i.e., $n=3$, for $%
\eta =0$ (Fig. 2(a)) and $\eta =\pi$, i.e., $k/l=1/2$ (Fig. 2(b)). Since $n$
is odd, the case in Fig. 2(a) is one of ordinary weak chaos (see Sec. IIIB).
On the other hand, in the case of Fig. 2(b), with $n=3$ and $l=2$, $%
l^{\prime}=2$ and the SWC condition (\ref{cE}) is thus satisfied for the
potential $V(x)=-\cos(x)$ ($N=1$). Indeed, the diffusion rate in this case
is clearly slower than that in Fig. 2(a). 
\begin{figure}[tbp]
\centering
\includegraphics[width=9cm,natwidth=850,natheight=400]{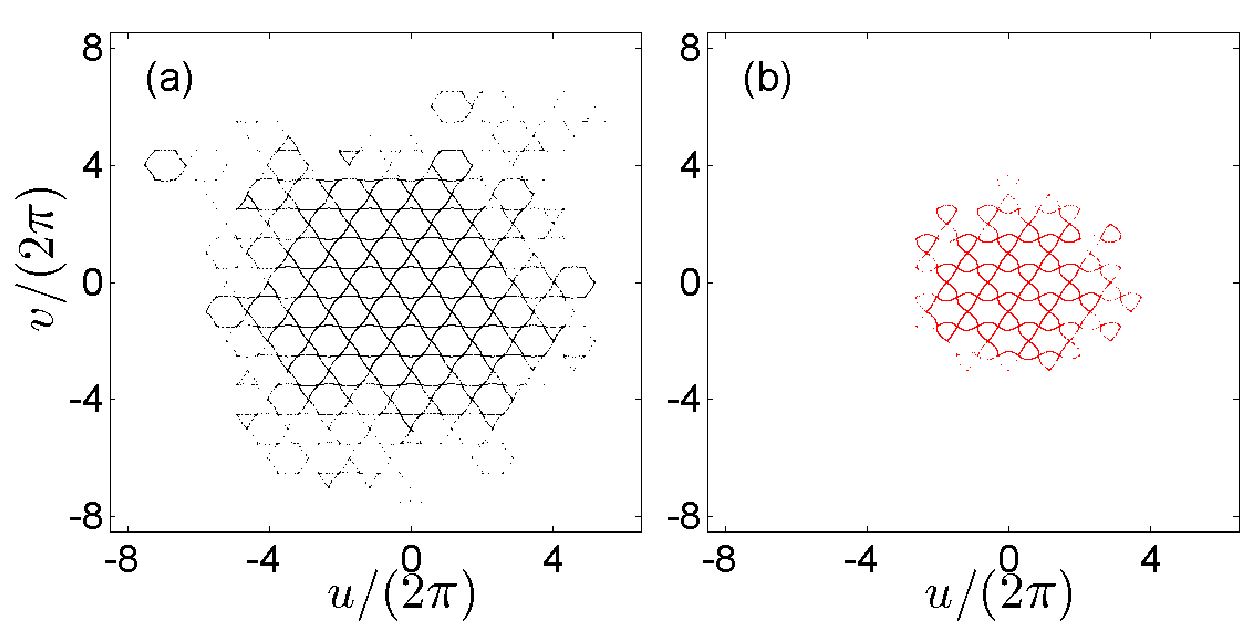}
\caption{(Color online) Similar to Fig. 1 with the only differences that the
number of iterations is 90000, $\protect\kappa =0.415$, $\protect\gamma =2%
\protect\pi /3$, and, in (b), $\protect\eta =\protect\pi$ and $x_{\mathrm{c}%
}^{(0)}=0$. Case (b) is a simple example of SWC for $\protect\eta\neq 0$,
when $x_{\mathrm{c}}$ is not a constant of the motion. In case (a), $\protect%
\eta =0$ and $x_{\mathrm{c}}=x_{\mathrm{c}}^{(0)}=0$, similarly to Fig. 1(a).
}
\label{fig2}
\end{figure}

\begin{center}
\textbf{D. SWC, linear instability, and chaotic layers}
\end{center}

Equation (\ref{SWC}) for SWC has straightforward implications for the linear
instability of periodic orbits of the map (\ref{Mh}). Each point of such an
orbit must generally be a fixed point of some iterate $s$ of $M_{\gamma
,\eta ,r}$. A hyperbolic fixed point of $M_{\gamma ,\eta ,r}^s$ is
characterized by its Lyapunov multiplier $\lambda$, i.e., the larger
eigenvalue ($>1$) of the linear-stability matrix $DM_{\gamma ,\eta , r}^s$
evaluated at the point. This matrix has unit determinant since $M_{\gamma
,\eta ,r}^s$ is area preserving. Using this fact and Eq. (\ref{SWC}), it is
easy to show that $\mathrm{Tr}(DM_{\gamma ,\eta ,r}^s)=2+O(\kappa^4)$, so
that $\lambda =1+O(\kappa ^4)$. Thus, the fixed point is significantly less
unstable than in the case of ordinary weak chaos [with $z_r=z_0+O(\kappa )$%
], where one has $\lambda =1+O(\kappa ^2)$. As a consequence, a SWC layer
emanating from an unstable fixed point should be narrower than an ordinary
weak-chaos layer.

This is illustrated in Figs. 3 and 4. The case in Fig. 3, corresponding to
the $E=0$ webs in Fig. 1, was analytically studied in Ref. \cite{prk} and
the following nonrigorous estimates of the chaotic-layer width $\Delta$ for $%
\kappa\ll 1$ were derived: $\Delta_0=(16\pi^3/\kappa)\exp(-\pi^2/\kappa )$
for $x_{\mathrm{c}}=0$ (weak chaos) and $\Delta_{\pi /2}=(4\pi^3/\kappa
^3)\exp(-\pi^2/\kappa ^2)$ for $x_{\mathrm{c}}=\pi /2$ (SWC). The main
dependence on $\kappa$ in these formulas is consistent with the fact that
the effective kick strength for SWC is $\kappa^2$ rather than $\kappa$;
thus, the ratio $\Delta_{\pi /2}/\Delta_0\rightarrow 0$ in the limit $%
\kappa\rightarrow 0$, as expected. Due to computational limitations, we
could not check numerically these formulas for sufficiently small $\kappa\ll
1$. Our accurate numerical results in Fig. 3 clearly show that the actual
value of $\Delta_{\pi /2}/\Delta_0$ for $\kappa =0.6$ is already very small,
in fact much smaller than its estimate from the formulas above. 
\begin{figure}[tbp]
\centering
\includegraphics[width=9cm,natwidth=650,natheight=500]{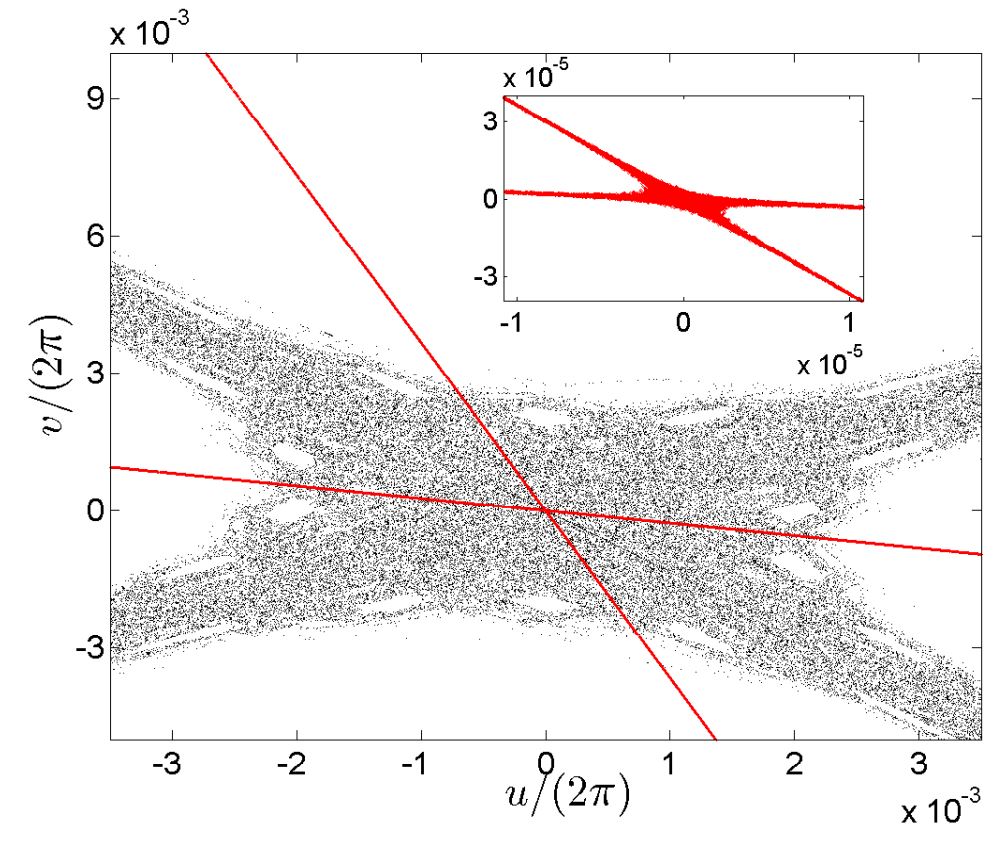}
\caption{(Color online) Chaotic layers around hyperbolic points of the webs
in Fig. 1. The large chaotic region (black dots) is the layer for the case
in Fig. 1(a) ($x_{\mathrm{c}}=0$). The two crossing lines correspond to the
much narrower SWC layer for the case in Fig. 1(b) ($x_{\mathrm{c}}=\protect%
\pi /2$). Only after a significant zoom, this chaotic layer becomes visible
(see inset). For the sake of comparison, the hyperbolic point in the latter
case was shifted so as to coincide with that in the former case.}
\label{fig3}
\end{figure}

The case in Fig. 4 corresponds to the webs in Fig. 2 for $E\neq 0$. We see
that the SWC layer [red (dark gray) region] is again narrower than the ordinary
weak-chaos one (black region). Due to computational limitations, we were not
able to get similar accurate plots for smaller values of $\kappa$ for which
the SWC layer is expected to be much narrower than the ordinary weak-chaos
one. 
\begin{figure}[tbp]
\centering
\includegraphics[width=8cm,natwidth=600,natheight=500]{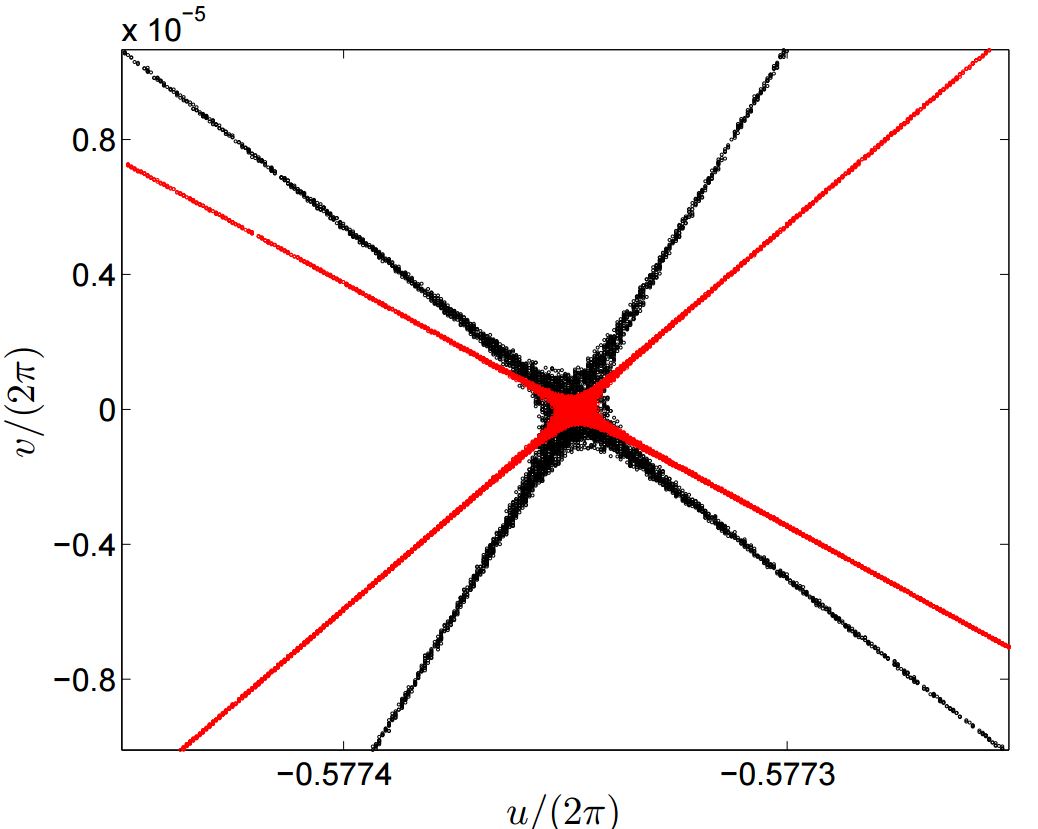}
\caption{(Color online) Similar to Fig. 3 but for the webs in Fig. 2. The
SWC chaotic layer [red (dark gray) region], corresponding to Fig. 2(b), is
again narrower than the ordinary weak-chaos one (black region) for the case
in Fig. 2(a).}
\label{fig4}
\end{figure}

\begin{center}
\textbf{IV. TRANSLATIONAL INVARIANCE, STOCHASTIC WEBS, AND EFFECTIVE
HAMILTONIANS}
\end{center}

We now assume that the basic map (\ref{cMhb}) has translational invariance
in the $(u,v)$ phase space. As shown in Appendix A, this is the case only
for the following values of $n$ in Eq. (\ref{rge}): $n=1,2,3,4,6$. As
already mentioned in Sec. IIIA, the map (\ref{cMhb}) for $n=1,2$ is
integrable. For $n=3,4,6$, the map may exhibit chaos emanating from
hyperbolic fixed points. Then, the translational invariance implies the
existence of an entire lattice of such points. Heteroclinic intersections of
the stable and unstable manifolds of neighboring points on this lattice may
generate global chaos on a stochastic web with triangular ($n=3$), square ($%
n=4$), or hexagonal ($n=6$) symmetry. Examples of both weak chaos and SWC on
stochastic webs for $n=4$ and $n=3$ are shown in Figs. 1 and 2,
respectively. Other examples of SWC webs for $n=4$ are shown in Fig. 5. 
\begin{figure}[tbp]
\includegraphics[width=8cm,natwidth=470,natheight=470]{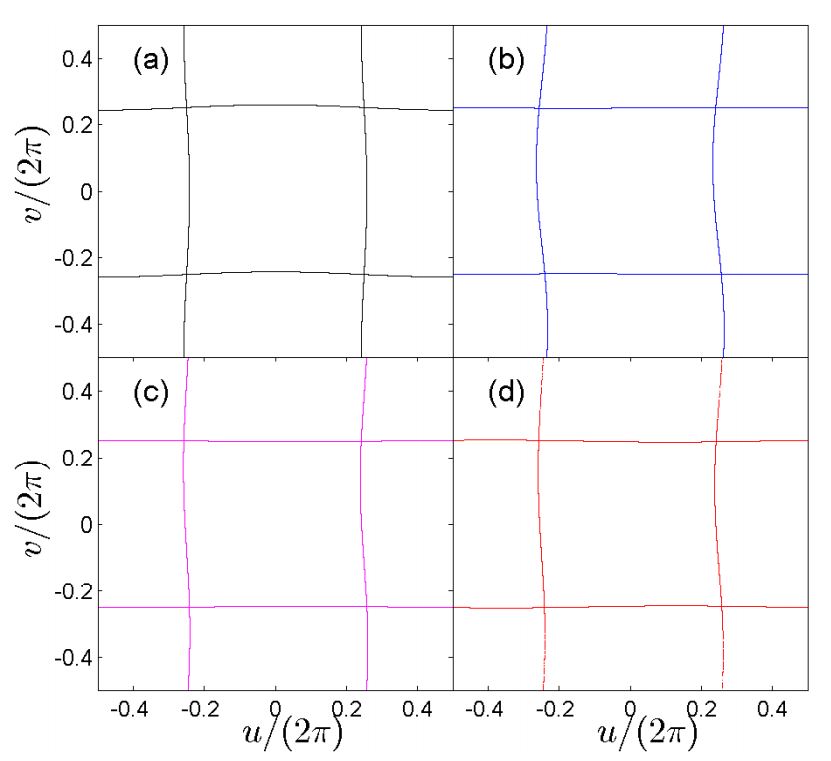}
\caption{(Color online) SWC stochastic webs (within the phase-space unit
cell of periodicity) generated by iterating a $20\times 20$ chaotic ensemble
with the map (\protect\ref{Mh}) for $\protect\kappa =0.1$, $f(x)=-\sin (x)$, 
$\protect\gamma =\protect\pi /2$ ($n=4$), and (a) $\protect\eta =0$, $x_{%
\mathrm{c}}^{(0)}=\protect\pi /2$; (b) $\protect\eta /(2\protect\pi ) =2/3$, 
$x_{\mathrm{c}}^{(0)}=0$; (c) $\protect\eta /(2\protect\pi ) =3/5$, $x_{%
\mathrm{c}}^{(0)}=0$; (d) $\protect\eta /(2\protect\pi ) =8/13$, $x_{\mathrm{%
c}}^{(0)}=0$. The values of $\protect\eta /(2\protect\pi )$ in (b)-(d)
correspond to rational approximants of the golden-mean inverse $(\protect%
\sqrt{5}-1)/2$. The plot for the $5/8$ approximant is not displayed. It
corresponds to a nongeneric case, see Sec. VB. The number of iterations is
120000 in (a) and 40000 in (b)-(d).}
\label{fig5}
\end{figure}

From now on, we shall restrict ourselves to the case of $n=4$ with the
standard potential $V(x)=-\cos (x)$, i.e., $N=1$ in Eq. (\ref{V}). Then, the
SWC condition (\ref{cE}) for $E\neq 0$ is: 
\begin{equation}  \label{cE1}
l^{\prime}>1,
\end{equation}
where $l^{\prime}$ is defined by $4/l=n^{\prime}/l^{\prime}$, with $%
(n^{\prime},l^{\prime})$ coprime integers. The basic map (\ref{cMhb})
reduces in our case to 
\begin{equation}  \label{cMh4}
M_{\gamma ,\eta ,r}:\ z_{r}=z_0-\kappa\sum_{j=0}^{r-1}\sin (x_{\mathrm{c}%
}^{(0)}+j\eta-v_j)e^{ij\pi /2},
\end{equation}
where $r=\mathrm{lcm}(4,l)=4l^{\prime}$ and, among the two values of $\gamma
=\pi /2$ or $3\pi /2$ for $n=4$, we choose $\gamma =\pi /2$ without loss of
generality (see note \cite{note1}). SWC for the map (\ref{cMh4}) is generic
under the condition (\ref{cE1}) in the sense that Eq. (\ref{SWC}) holds for
arbitrary value of $x_{\mathrm{c}}^{(0)}$. Also, a SWC stochastic web will
emerge for all $l^{\prime}>2$ and for all $x_{\mathrm{c}}^{(0)}$, see Sec.
IVB; examples of such generic SWC webs are given in Figs. 5(b)-5(d). In Sec.
IVA, we present numerical results for the global SWC diffusion on stochastic
webs. In Sec. IVB, we derive effective Hamiltonians giving the SWC web
skeleton for $l^{\prime}>2$.

\begin{center}
\textbf{A. Numerical results for global SWC diffusion}
\end{center}

Global chaos on stochastic webs under the basic map (\ref{cMh4}) is
illustrated in Figs. 1 and 5 for different values of $\eta/(2\pi )$. One
expects this chaotic motion to exhibit a normal or anomalous diffusive
behavior: 
\begin{equation}  \label{D}
\langle |z_{rs}-z_0|^2\rangle _{\mathcal{E}} \approx 2D s^{\mu},
\end{equation}
where $\langle \ \rangle _{\mathcal{E}}$ denotes average over an ensemble $%
\mathcal{E}$ of initial conditions $z_0=u_0+iv_0$ within the chaotic layer, $%
r=4l^{\prime}$, $D$ is the diffusion coefficient, and $\mu$ is the diffusion
exponent; $\mu =1$ ($\mu \neq 1$) corresponds to normal (anomalous)
diffusion. Figure 6 shows $\langle |z_{rs}-z_0|^2\rangle _{\mathcal{E}}$
versus $rs$ at fixed $r$ for $\kappa =0.1$ and different values of $\eta$
and $x_{\mathrm{c}}^{(0)}$. Clearly, the SWC diffusion for $\eta =0$ and $x_{%
\mathrm{c}}^{(0)}=\pi /2$ or for $\eta\neq 0$ and $l^{\prime}>1$ is
significantly suppressed relative to the weak-chaos one for $\eta =0$ and $%
x_{\mathrm{c}}^{(0)}=0$. The results in Fig. 6 appear in Fig. 7 as log-log
plots, showing that within the large time interval considered the diffusion
is anomalous, i.e., it is a subdiffusion ($\mu <1$) with $\mu$ ranging
between $\approx 0.5$ and $\approx 0.6$. Such a subdiffusion is
theoretically expected \cite{y} to be a transient behavior followed, at
sufficiently large times, by normal diffusion ($\mu =1$). We were able to
observe a transition to normal diffusion only for $\kappa > 0.2$. Examples
of such a transition are shown in Fig. 8. 
\begin{figure}[tbp]
\centering
\includegraphics[width=8.5cm,natwidth=600,natheight=500]{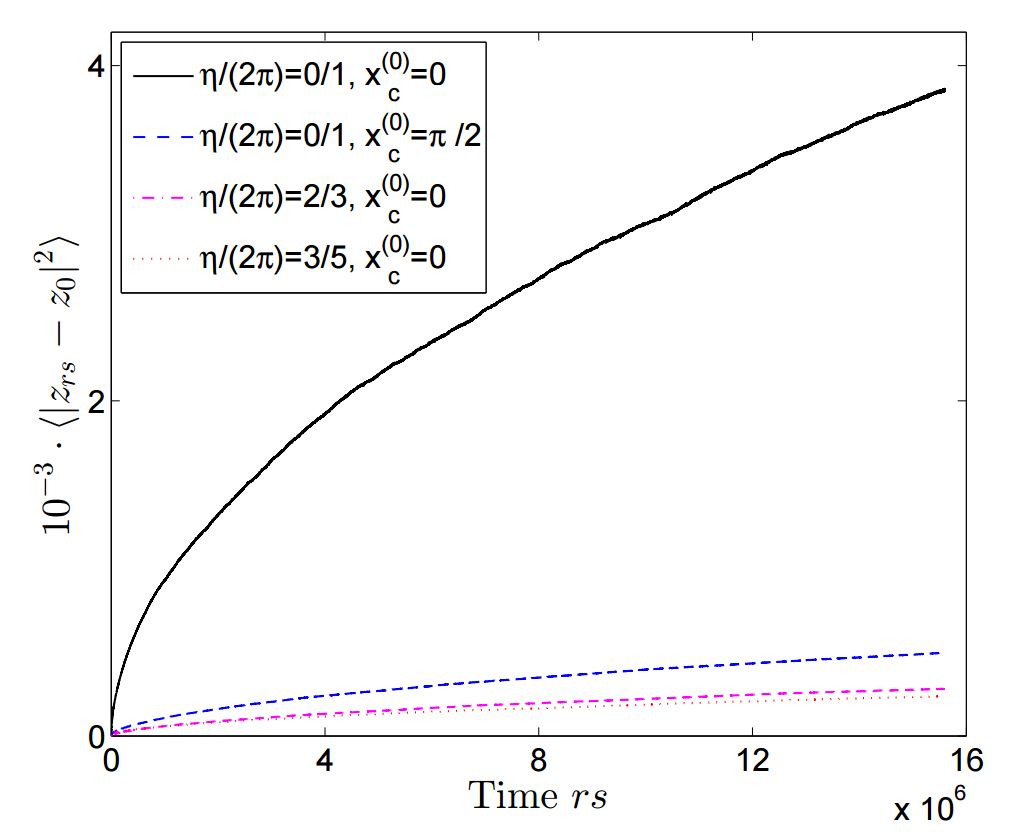}
\caption{(Color online) Diffusive behavior (\protect\ref{D}) of a $100\times
100$ chaotic ensemble after $s$ iterations of the map (\protect\ref{cMh4})
for $\protect\kappa =0.1$, maximal time interval $rs_{\mathrm{max}%
}=1.56\cdot 10^7$, and different values of $\protect\eta$ and $x_{\mathrm{c}%
}^{(0)}$ specified in the legend in order of descending curves. One has: $r=4
$, $s_{\mathrm{max}}=3.9\cdot 10^6$ for $\protect\eta =0$; $r=12$, $s_{%
\mathrm{max}}=1.3\cdot 10^6$ for $\protect\eta /(2\protect\pi )=2/3$; $r=20$%
, $s_{\mathrm{max}}=7.8\cdot 10^5$ for $\protect\eta /(2\protect\pi )=3/5$.}
\label{fig6}
\end{figure}
\begin{figure}[tbp]
\centering
\includegraphics[width=8.5cm,natwidth=600,natheight=500]{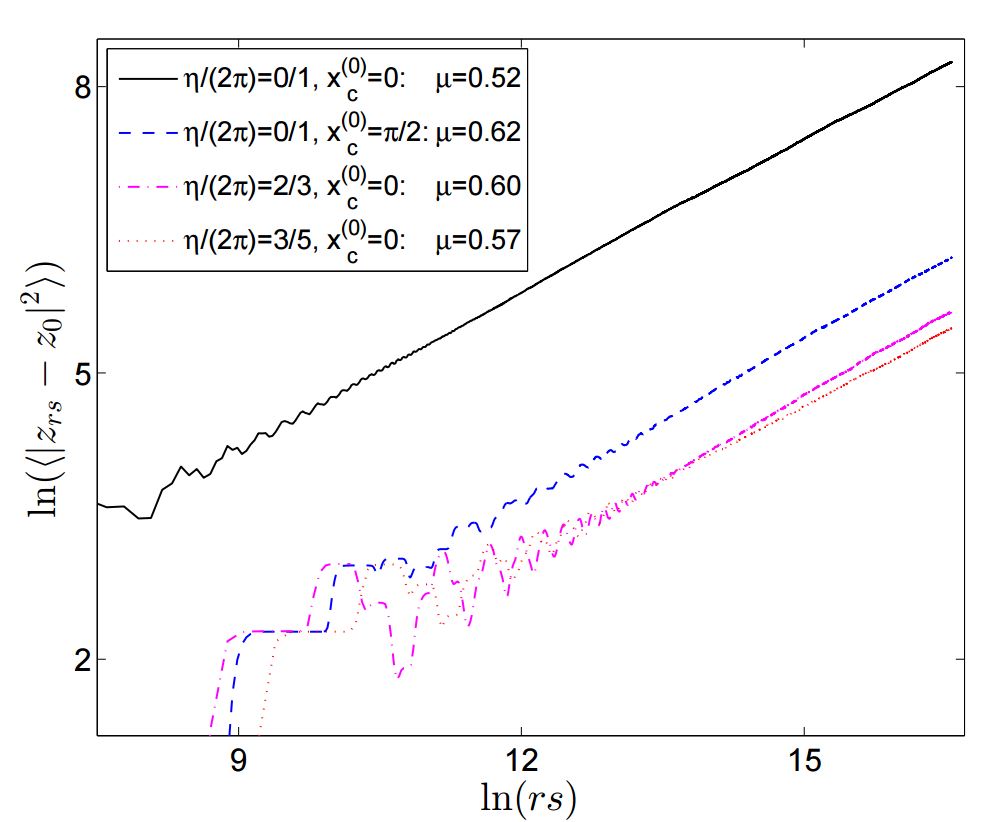}
\caption{(Color online) Log-log plots of the curves in Fig. 6. The slope of
each plot is approximately the anomalous diffusion exponent $\protect\mu$
(given in the legend) over the time interval considered.}
\label{fig7}
\end{figure}
\begin{figure}[tbp]
\centering
\includegraphics[width=8.5cm,natwidth=600,natheight=500]{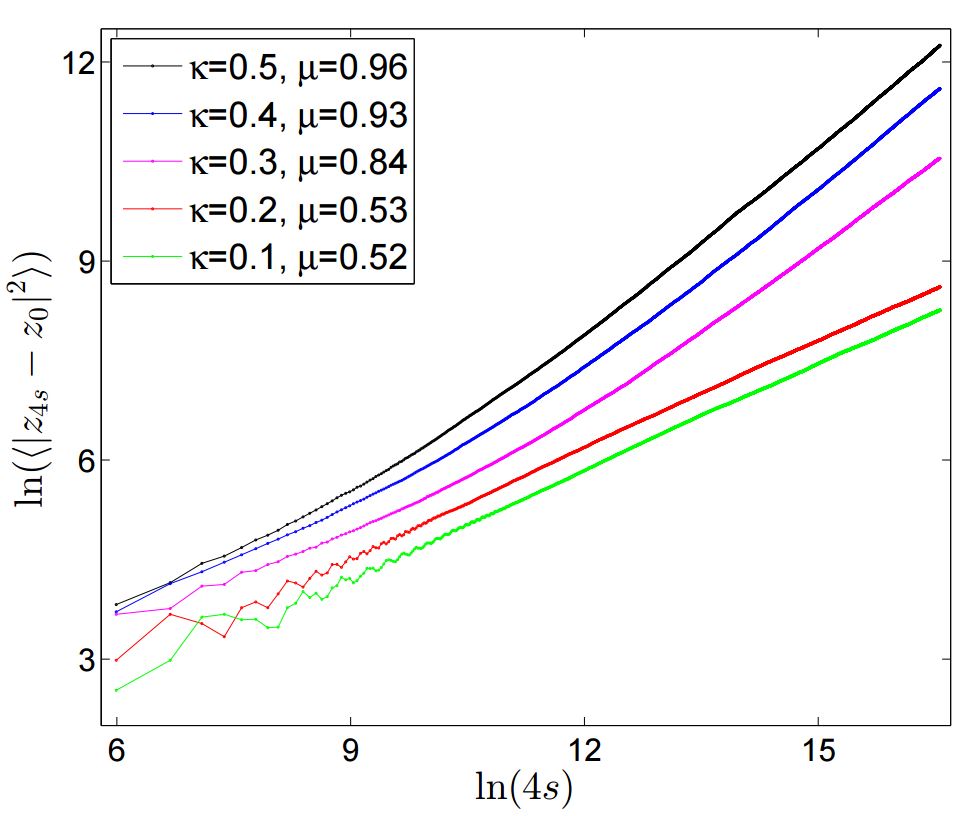}
\caption{(Color online) Log-log plots of the diffusive behavior (\protect\ref%
{D}) of a $100\times 100$ chaotic ensemble after $s$ iterations of the map (%
\protect\ref{cMh4}) for $\protect\eta =0$ ($r=4$), maximal time interval $%
4s_{\mathrm{max}}=1.56\cdot 10^7$, $x_{\mathrm{c}}^{(0)}=0$, and different
values of $\protect\kappa$ specified in the legend in order of descending
curves. The anomalous diffusion exponent $\protect\mu$ given in the legend
is the slope of the corresponding curve near $s=s_{\mathrm{max}}$. For $%
\protect\kappa > 0.2$, the slope appears to increase toward an asymptotic
value of $\protect\mu =1$, indicating a transition to normal diffusion at
sufficiently large times.}
\label{fig8}
\end{figure}
\newpage

\begin{center}
\textbf{B. Integrable effective Hamiltonians for SWC on stochastic webs}
\end{center}

We show here that for sufficiently small $\kappa$, i.e., $r\kappa\ll 1$ ($%
r=4l^{\prime}$), the basic map (\ref{cMh4}) can be approximately replaced by
the Hamilton equations with an integrable effective Hamiltonian $H_{\mathrm{e%
}}$. Considering $r\kappa$ as a small time step $\Delta t$, we replace $%
(u_r-u_0)/(r\kappa )$ and $(v_r-v_0)/(r\kappa )$ by the time derivative $%
\dot{u}$ and $\dot{v}$, respectively. We then write the real and imaginary
parts of Eq. (\ref{cMh4}) as approximate Hamilton equations 
\begin{equation}  \label{hes}
\dot{u}\approx \left.\frac{\partial H_{\mathrm{e}}}{\partial v}%
\right|_{u_0,v_0} =\Re (F),\ \ \ \dot{v}\approx -\left.\frac{\partial H_{%
\mathrm{e}}}{\partial u}\right|_{u_0,v_0}=\Im (F),
\end{equation}
where 
\begin{equation}  \label{F}
F=-\frac{1}{r}\sum_{j=0}^{r-1}\sin(x_{\mathrm{c}}^{(0)}+j\eta-v_j)e^{ij\pi
/2}.
\end{equation}
In the SWC case of $l^{\prime}>1$ [Eq. (\ref{cE1})], $F$ must be of order $%
O(\kappa )$ because of Eq. (\ref{SWC}). Indeed, after a lengthy but
straightforward calculation given in Appendix B, we find that the
leading-order term of $F$ in a power expansion in $\kappa$ is, for $%
l^{\prime}>2$, 
\begin{equation}  \label{Feg}
F=\frac{\kappa}{8\cos(\eta )}[\cos(u_0)\sin(v_0)-i\sin(u_0)\cos(v_0)],
\end{equation}
independent of $x_{\mathrm{c}}^{(0)}$. For $l^{\prime}=2$, on the other
hand, we show in Appendix B that $F$ depends on $x_{\mathrm{c}}^{(0)}$. The
implications of this dependence are studied in detail in Sec. VB. Here, we
consider the more general case of $l^{\prime}>2$, with $F$ given by Eq. (\ref%
{Feg}). Using Eq. (\ref{Feg}) in Eqs. (\ref{hes}), we obtain the
leading-order term of the SWC effective Hamiltonian for $l^{\prime}>2$, up
to an additive constant: 
\begin{equation}  \label{He}
H_{\mathrm{e}}= -\frac{\kappa}{8\cos(\eta)}\cos(u_0)\cos(v_0).
\end{equation}
The integrable Hamiltonian (\ref{He}) gives approximations to the orbits of
the map (\ref{cMh4}) as ``level sets" $H_{\mathrm{e}}=C$ for constant
``energy" $C$, $|C|\leq\kappa/[8\cos(\eta)]$. As shown in Appendix C, a
stochastic web must correspond to the level set $H_{\mathrm{e}}=0$. From Eq.
(\ref{He}), this level set is the union of the straight lines $%
u_0=\pi/2+a_1\pi$ and $v_0=\pi/2+a_2\pi$, for all integers $(a_1,a_2)$. This
set, a grid with a $\pi\times\pi$ unit cell, gives the integrable skeleton
of the stochastic web in the limit of $\kappa\rightarrow 0$. Stochastic webs
close to this skeleton are shown in Fig. 5; the case of $\eta =0$ and $x_{%
\mathrm{c}}^{(0)}=\pi /2$ in Fig. 5(a) also exhibits this skeleton, see Ref. 
\cite{prk}.

\begin{center}
\textbf{V. BALLISTIC MOTION IN WEAK-CHAOS AND SWC CASES}
\end{center}

In Sec. IVB, we have derived effective Hamiltonians for the basic map (\ref%
{cMh4}) in the generic SWC case of $l^{\prime}>2$, where a stochastic web
always exists for \emph{all} $x_{\mathrm{c}}^{(0)}$. Here we consider the
exceptional cases of $l^{\prime}=1$ [ordinary weak chaos, since the SWC
condition (\ref{cE1}) is not satisfied] and $l^{\prime}=2$ (SWC). We show
that in these cases ballistic motion arises for almost all values of $x_{%
\mathrm{c}}^{(0)}$ while stochastic webs emerge only in very small intervals
of $x_{\mathrm{c}}^{(0)}$.

\begin{center}
\textbf{A. Weak-chaos case of $l^{\prime}=1$}
\end{center}

Among the different values of $\eta$ corresponding to $l^{\prime}=1$, it is
sufficient to consider $\eta=\pi/2$, without loss of generality (see note 
\cite{note2}). To derive an effective Hamiltonian in this case, let us first
calculate the map (\ref{cMh4}) for $l^{\prime}=1$ and $\eta=\pi/2$ to first
order in $\kappa$; to this end, it is sufficient to write $v_j$ to zero
order in $\kappa$: $v_1=-u_0$, $v_2=-v_0$, and $v_3=u_0$ [from Eqs. (\ref{Mh}%
) for $\kappa =0$]. Then, proceeding as in Sec. IVB, we find that Eqs. (\ref%
{hes}) are satisfied with $F=(z_4-z_0)/(4\kappa)\approx \dot{z}$ explicitly
given by: 
\begin{equation}  \label{Fa}
F=-\frac{1}{2}\left[\sin(x_{\mathrm{c}}^{(0)})\cos(v_0)+i\cos(x_{\mathrm{c}%
}^{(0)})\cos(u_0)\right]+O(\kappa).
\end{equation}
Using Eq. (\ref{Fa}) in Eqs. (\ref{hes}) and integrating, we get the
leading-order term of $H_{\mathrm{e}}$ up to an additive constant: 
\begin{equation}  \label{Hea}
H_{\mathrm{e}}= \frac{1}{2}\left[\cos(x_{\mathrm{c}}^{(0)})\sin(u_0)-\sin(x_{%
\mathrm{c}}^{(0)})\sin(v_0)\right].
\end{equation}

For a general level set $H_{\mathrm{e}}=C$, we get from Eq. (\ref{Hea}): 
\begin{equation}
\sin (u_{0})=\tan (x_{c}^{(0)})\sin (v_{0}))+2C/\cos (x_{c}^{(0)}).
\label{ss}
\end{equation}%
We see from Eq. (\ref{ss}) that for $|\tan (x_{c}^{(0)})|<1$ and any $C$, $%
u_{0}$ will cover only part of the interval $[0,2\pi )$ when $v_{0}$ varies
in this interval. This case corresponds to ballistic motion in the $v$
direction, i.e., $v_{s}$ increases, on the average, linearly in
\textquotedblleft time" $s$; see, e.g., Fig. 9(a). For $|\tan
(x_{c}^{(0)})|>1$ and any $C$, $v_{0}$ will cover only part of $[0,2\pi )$
when $u_{0}$ varies in $[0,2\pi )$. This case corresponds to ballistic
motion in the $u$ direction. Only in the case of $|\tan (x_{c}^{(0)})|=1$
and $C=0$, both $u_{0}$ and $v_{0}$ will cover all the interval $[0,2\pi )$.
As shown in Appendix C, this is the only case where a stochastic web can
emerge in the framework of the effective Hamiltonian (\ref{Hea}). An example
of stochastic web for the critical value of $x_{c}^{(0)}=\pi /4$ [$\tan
(x_{c}^{(0)})=1$] is shown in Fig. 9(b). It agrees very well with the web
skeleton which, from Eq. (\ref{Hea}) with $H_{\mathrm{e}}=0$ and $%
x_{c}^{(0)}=\pi /4$, is the union of the lines $u_{0}-v_{0}=2a_{1}\pi $ and $%
u_{0}+v_{0}=(2a_{2}+1)\pi $ for all integers $(a_{1},a_{2})$. Because of the
effective-Hamiltonian approximation and the small but finite width of the
chaotic layer, a stochastic web will actually exists in small intervals of $%
x_{c}^{(0)}$ around the critical values of $x_{c}^{(0)}$ satisfying $|\tan
(x_{c}^{(0)})|=1$. 
\begin{figure}[tbp]
\centering
\includegraphics[width=8.5cm,natwidth=600,natheight=300]{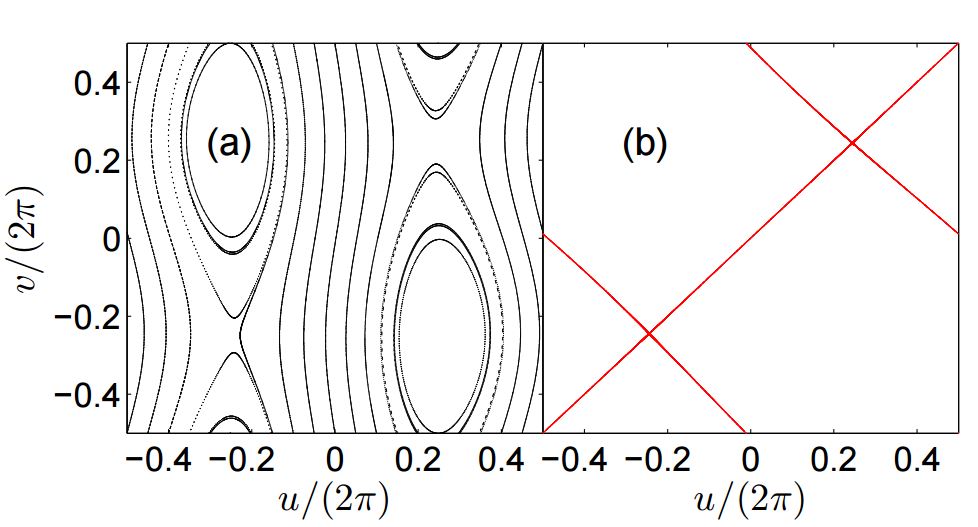}
\caption{(Color online) Orbits of the map (\protect\ref{cMh4}) within the
phase-space unit cell of periodicity for $\protect\kappa =0.1$, $\protect%
\eta =\protect\pi /2$, and two values of $x_{c}^{(0)}$: (a) $x_{c}^{(0)}=%
\protect\pi /8$, corresponding to ballistic orbits in the $v$ direction; (b) 
$x_{c}^{(0)}=\protect\pi /4$, the critical value of $x_{c}^{(0)}$ for the
emergence of a stochastic web.}
\label{fig9}
\end{figure}

\begin{center}
\textbf{B. SWC case of $l^{\prime}=2$}
\end{center}

By considerations similar to those in note \cite{note2}, we find that among
the different values of $\eta$ corresponding to $l^{\prime}=2$, it is
sufficient to consider $\eta=\pi/4$ and $\eta=3\pi/4$, without loss of
generality. We shall restrict ourselves here to the case of $\eta=\pi/4$
since the case of $\eta=3\pi/4$ can be treated in a very similar way. The
expression (\ref{F}) for $l^{\prime}=2$ and $\eta=\pi/4$ is calculated in
Appendix B to first order in $\kappa$ and is given by Eq. (\ref{Feg2})
there. Using this expression in Eqs. (\ref{hes}) and integrating, we get the
leading-order term of $H_{\mathrm{e}}$ up to an additive constant: 
\begin{eqnarray}
H_{\mathrm{e}} &=& -\frac{\sqrt{2}\kappa}{8} [\cos(u_0)\cos(v_0)  \notag \\
&+&\cos(2x_{\mathrm{c}}^{(0)})\sin(u_0)\cos(v_0)  \notag \\
&-& \sin(2x_{\mathrm{c}}^{(0)})\cos(u_0)\sin(v_0)].  \label{Heb}
\end{eqnarray}
We show in Appendix C that a stochastic web must again correspond to the
level set $H_{\mathrm{e}}=0$ of Eq. (\ref{Heb}). To determine the critical
values of $x_{\mathrm{c}}^{(0)}$ for this web, let us write Eq. (\ref{Heb})
for $H_{\mathrm{e}}=0$ as follows: 
\begin{equation}  \label{bal2}
1+\cos(2x_{\mathrm{c}}^{(0)})\tan(u_0)=\sin(2x_{\mathrm{c}}^{(0)})\tan(v_0),
\end{equation}
where we assumed that both $\cos(2x_{\mathrm{c}}^{(0)})$ and $\sin(2x_{%
\mathrm{c}}^{(0)})$ are nonzero, i.e., $x_{\mathrm{c}}^{(0)}\neq a\pi/4$ for
integer $a$. Then, by Eq. (\ref{bal2}), $v_0$ is a monotonically increasing
function of $u_0$ in the interval $[-\pi /2,\pi /2]$. This corresponds to
``diagonal" ballistic orbits for which both $u_s$ and $v_s$ increase, on the
average, linearly in ``time" $s$ [unlike the ``horizontal" or "vertical"
ballistic orbits in the case of Sec. VA, see Fig. 9(a)]. Consider now the
special values above of $x_{\mathrm{c}}^{(0)}$, $x_{\mathrm{c}}^{(0)}= a\pi/4
$, $a$ integer. For example, in the case of $x_{\mathrm{c}}^{(0)}= \pi/4$,
Eq. (\ref{Heb}) for $H_{\mathrm{e}}=0$ gives 
\begin{equation}  \label{ccs}
\cos(u_0)[\cos(v_0)-\sin(v_0)]=0  \notag
\end{equation}
with solutions $u_0=\pi/2+a_1\pi$ and $v_0=\pi/4+a_2\pi$ for all integers $%
(a_1,a_2)$. These vertical and horizontal lines define a web skeleton
similar to that in the generic SWC case of $l^{\prime}>2$ in Sec. IVB. The
positions of the lines modulo $\pi$ for the four values of $x_{\mathrm{c}%
}^{(0)}= a\pi/4$ modulo $\pi$ are summarized in Table \ref{table1}. 
\begin{table}[h]
\caption{Straight lines defining the web skeleton for $\protect\eta=\protect%
\pi/4$.}
\label{table1}\centering
\begin{tabular}{|c|c|c|}
\hline
$x_{\mathrm{c}}^{(0)}$\ mod($\pi$) & $u_0$\ mod($\pi$) & $v_0$\ mod($\pi$)
\\ \hline
$0$ & $3\pi/4$ & $\pi/2$ \\ \hline
$\pi/4$ & $\pi/2$ & $\pi/4$ \\ \hline
$\pi/2$ & $\pi/4$ & $\pi/2$ \\ \hline
$3\pi/4$ & $\pi/2$ & $3\pi/4$ \\ \hline
\end{tabular}%
\end{table}

\begin{center}
\textbf{C. Transition from ballistic motion to a stochastic web}
\end{center}

As shown above, the exceptional cases of $l^{\prime }=1,2$ feature ballistic
motion for almost all $x_{\mathrm{c}}^{(0)}$ and stochastic webs in small
intervals of $x_{\mathrm{c}}^{(0)}$ around critical values $x_{\mathrm{c}%
}^{(0)}=x_{\mathrm{cc}}^{(0)}$. As $x_{\mathrm{c}}^{(0)}$ approaches some
value $x_{\mathrm{cc}}^{(0)}$, ballistic motion should change gradually to
transport not faster than chaotic diffusion on the web. Therefore, the
average ballistic velocity is expected to vanish as $x_{\mathrm{c}%
}^{(0)}\rightarrow x_{\mathrm{cc}}^{(0)}$. Defining the average velocity in,
say, the $v$ direction by 
\begin{equation}
I_{v}\equiv \lim_{s\rightarrow \infty }\frac{1}{rs}\left\langle
|v_{rs}-v_{0}|\right\rangle _{\mathcal{E}},  \label{Iv}
\end{equation}%
for some suitable ensemble $\mathcal{E}$ of initial conditions, our
numerical results indicate that $I_{v}$ vanishes almost linearly as $x_{%
\mathrm{c}}^{(0)}\rightarrow x_{\mathrm{cc}}^{(0)}$. See, e.g., Fig. 10 for
the case in Fig. 9, with $x_{\mathrm{cc}}^{(0)}=\pi /4$. 
\begin{figure}[tbp]
\centering
\includegraphics[width=8.5cm,natwidth=600,natheight=500]{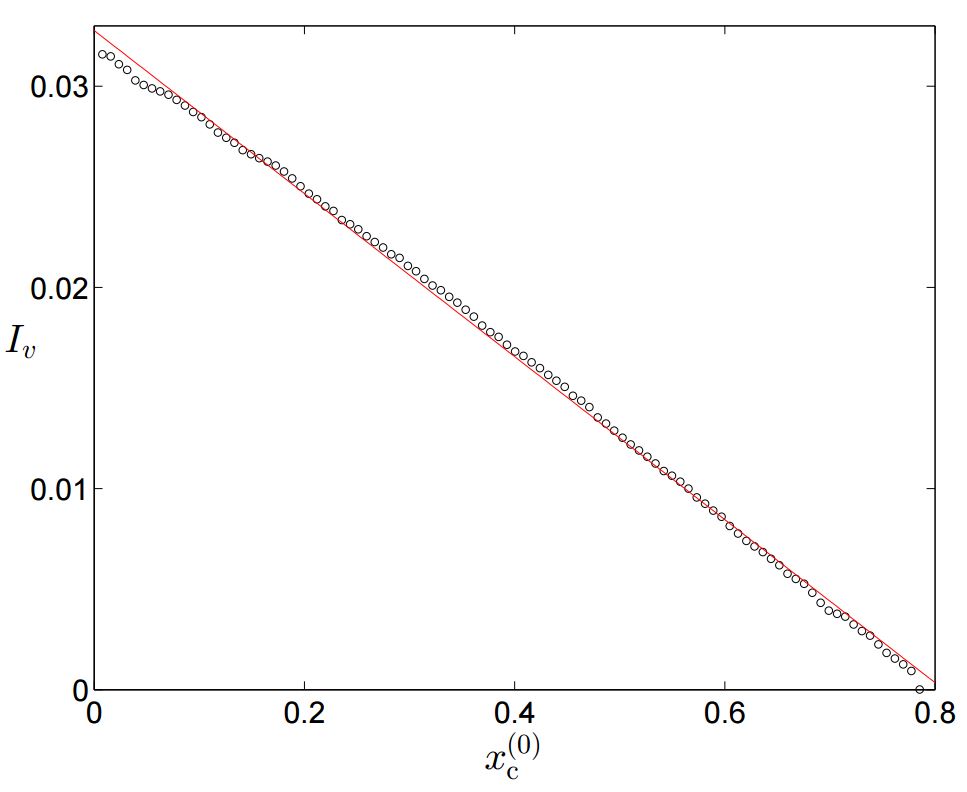}
\caption{(Color online) The average velocity (\protect\ref{Iv}) versus $x_{%
\mathrm{c}}^{(0)}$ for the case in Fig. 9 [$\protect\kappa =0.1$, $\protect%
\eta =\protect\pi /2$ ($r=4$), and $x_{\mathrm{cc}}^{(0)}=\protect\pi /4$]; $%
I_{v}$ was calculated using Eq. (\protect\ref{Iv}) with $s=30000$ and an
ensemble $\mathcal{E}$ of $20\times 20$ initial conditions in the $2\protect%
\pi \times 2\protect\pi $ unit cell of periodicity.}
\label{fig10}
\end{figure}

{\ }\newline

\begin{center}
\textbf{VI. SUMMARY AND CONCLUSIONS}
\end{center}

In this paper, we have introduced a realistic non-KAM Hamiltonian system,
the KHS defined by Eq. (\ref{H}) or by Eqs. (\ref{eKHO}) and (\ref{xct}),
and we have performed a first study of its classical dynamics and transport
in the weak-chaos regime of small nonintegrability parameter $\kappa
=K/\omega$. We have shown that the classical Hall effect from the
perpendicular magnetic ($\mathbf{B}$) and electric ($\mathbf{E}$) fields is
relatively stabilizing: Under resonance conditions (\ref{rge}), it induces a
suppression of the weak chaos into superweak chaos (SWC) [defined by Eq. (%
\ref{SWC})] for a generic family of periodic kicking potentials (\ref{V}).
The SWC, either local or global, manifests itself in a decrease of the
instability of periodic orbits and in a narrowing of the corresponding
chaotic layers relative to the ordinary weak-chaos case, see Figs. 3 and 4.
Also, for global SWC on stochastic webs, the chaotic diffusion on the web is
much slower than the weak-chaos one, see Figs. 6 and 7. This global SWC was
studied in detail in Secs. IV and V in the case of $\gamma =\pi /2$ and the
standard potential $V(x)=-\cos (x)$. We have shown that global SWC on
stochastic webs is a generic phenomenon, occurring for all $l^{\prime}>2$
[where $l^{\prime}$ is defined by $4/l=n/l^{\prime}$ with $(n,l^{\prime})$
coprime integers] and for all initial values $x_{\mathrm{c}}^{(0)}$ of $x_{%
\mathrm{c}}$. In the special cases of $l^{\prime}=1$ and $l^{\prime}=2$, one
has, respectively, weak-chaos and SWC ballistic motion for almost all values
of $x_{\mathrm{c}}^{(0)}$. A transition from ballistic motion to
stochastic-web diffusion occurs when $x_{\mathrm{c}}^{(0)}$ approaches some
critical values $x_{\mathrm{cc}}^{(0)}$. The results in Secs. IV and V may
be generalized to arbitrary potential (\ref{V}).

The relatively slow SWC diffusion on stochastic webs in the area-preserving
map (\ref{Mh}) is apparently the two-dimensional closest analog to the
Arnol'd web diffusion in higher dimensional systems. However, while the rate
of the latter diffusion decreases exponentially with the nonintegrability
parameter \cite{a1,akn,kl}, the SWC-diffusion rate is expected to decrease
only algebraically with this parameter.

Maps for kicked systems are usually derived by integrating over the delta
functions in time in Hamilton equations. The map variables cannot be defined
precisely at the kicks (the delta functions) but only in an infinitesimal
vicinity of them, e.g, at the times $sT-0$ in the map (\ref{Mh}). By
considering the periodic delta function in Eq. (\ref{H0}) as a limit of a
continuous periodic function with a broad spectrum and by using a method of
canonical transformation of variables \cite{sa1}, one can construct a
``canonical" web map \cite{sa2} for the system (\ref{H0}) (without electric
field, $\eta =0$) with variables defined precisely at the kicking times $sT$%
. As one could expect, this map was found to differ significantly from the
ordinary web map [Eq. (\ref{Mh}) for $\eta =0$] defined at $t=sT-0$. In
particular, the stochastic-web width for the canonical web map can be much
smaller than that for the ordinary one. It would be interesting to
investigate to what extent this phenomenon is similar to the SWC, as defined
in Sec. IIIA, and to study the KHS ($\eta\neq 0$) using the canonical-map
approach. It was also found \cite{sa2} that the stochastic-web width for the
ordinary web map ($\eta =0$) can be quite small for $n=3$ and kicking
potential $V(x)=-\cos (x)$. Since this phenomenon occurs for odd $n$ and
even $V(x)$, opposite to the SWC conditions for $\eta =0$ in Sec. IIIB, it
cannot be identified with SWC and therefore needs a separate detailed study.

The results in this paper should form the basis for the study of
quantum-chaos phenomena in the quantized KHS. It is known \cite{d,dd} that a
quantum manifestation of SWC for $E=0$ [system (\ref{H0})] is quantum
antiresonance, i.e., the evolution operator for some values of a scaled
Planck constant is identically a phase factor, so that no wave-packet moves.
Since SWC for $E=0$ is rare, i.e., it occurs under nongeneric conditions
(see Secs. I and IIIB), also quantum antiresonance for $E=0$ is a rare
phenomenon. For $E\neq 0$, on the other hand, quantum antiresonance is
expected to emerge under generic conditions similar to those for SWC, see
Sec. IIIC. The validity of this expectation and other phenomena in the
quantized KHS are planned to be investigated in future works. Finally, we
remark that since the KHS is essentially equivalent to a modulated kicked
harmonic oscillator [see Eq. (\ref{eKHO}) with Eq. (\ref{xct})], the
quantized KHS should be experimentally realizable using atom-optics methods
like the ordinary quantum kicked harmonic oscillator \cite{dmcw}.

\begin{center}
\textbf{APPENDIX A}
\end{center}

Consider a lattice $a_1Z_1+a_2Z_2$ in the complex plane, where $(a_1,a_2)$
are all integers and $(Z_1,Z_2)$ are basic lattice vectors. We show here
that the map (\ref{cMhb}), for all $\kappa$, is invariant in $z$ under
translations on this lattice only if $n=1,2,3,4,6$ in Eq. (\ref{rge}). Since
the sum in Eq. (\ref{cMhb}) involves all the iterates from $j=0$ to $j=r-1$,
this invariance is generally possible only if the lattice $a_1Z_1+a_2Z_2$ is
invariant under the one-iteration map (\ref{cMh}). Now, in the limit of $%
\kappa\rightarrow 0$, the map (\ref{cMh}) and its inverse are just rotations 
$\exp (\pm i\gamma)$ by angles $\pm\gamma$. The invariance of $a_1Z_1+a_2Z_2$
under $\exp (\pm i\gamma)$ implies, in particular, that the vector 
\begin{equation}  \label{pmgz1}
(e^{i\gamma}+e^{-i\gamma})Z_1=2\cos (\gamma )Z_1
\end{equation}
belongs to the lattice above. Therefore, one must have $2\cos (\gamma )=a_1$
(integer). The only solutions of the latter equation are precisely the
values of $\gamma$ in Eq. (\ref{rge}) with $n=1,2,3,4,6$.

We also determine explicitly here basic lattice vectors $Z_1$ and $Z_2$ for
the values above of $n$. Since the function $f(x)$ in Eq. (\ref{cMh}) is $%
2\pi$-periodic, it is easy to see that for $n=1,2$ ($\gamma =0,\pi$), one
can choose $Z_1$ to be an arbitrary real number and $Z_2=2\pi i$; for $n=4$ (%
$\gamma =\pi /2,3\pi /4$), one can choose $Z_1=2\pi$ and $Z_2=2\pi i$. In
the triangular case of $n=3$ (e.g., $\gamma =2\pi /3$), let us write $%
Z_{1,2}=\bar{u}+i\bar{v}$. Then, the invariance of the second equation of
the map (\ref{Mh}) under a translation by $Z_{1,2}$ implies that 
\begin{equation}  \label{inv3}
\bar{v}=2\pi a,\ \ \ (\sqrt{3}\bar{u}+\bar{v})/2=2\pi\bar{a},
\end{equation}
where $(a,\bar{a})$ are integers. Two independent pairs of minimal values $%
(a,\bar{a})$ are $(1,0)$ and $(0,1)$. The corresponding pairs $(\bar{u},\bar{%
v})$ from Eq. (\ref{inv3}) give the basic lattice vectors $Z_1=2\pi (1/\sqrt{%
3}+i))$ and $Z_2=4\pi/\sqrt{3}$, both of length $4\pi/\sqrt{3}$ and defining
a unit cell of area $8\pi ^2/\sqrt{3}$. Similar results are obtained in the
hexagonal case of $n=6$.

\begin{center}
\textbf{APPENDIX B}
\end{center}

We calculate here $F$ in Eq. (\ref{F}) to first order in $\kappa$ under the
SWC condition $l^{\prime}>1$. In particular, the expression (\ref{Feg}) for $%
l^{\prime}>2$ is derived.

Since $r=4l^{\prime}$, we write in Eq. (\ref{F}) $j=4b+w$, where $b=0,\dots
,l^{\prime}-1$ and $w=0,\dots ,3$: 
\begin{equation}  \label{Fe}
F=-\frac{1}{r}\sum_{w=0}^{3}e^{iw\pi /2}\sum_{b=0}^{l^{\prime}-1}\sin[x_{%
\mathrm{c}}^{(0)}+(4b+w)\eta-v_{4b+w}].
\end{equation}
Let us calculate $v_{4b+w}$ in Eq. (\ref{Fe}) to first order in $\kappa$.
The map (\ref{Mh}) for $f(x)=-\sin(x)$ and $\gamma =\pi /2$ is 
\begin{equation}  \label{Mh4}
u_{s+1}=v_s,\ \ \ v_{s+1}=-u_s+\kappa\sin(x_{\mathrm{c}}^{(0)}+s\eta -v_s).
\end{equation}
By iterating Eqs. (\ref{Mh4}), we find that 
\begin{eqnarray}
v_{s+4}=v_s&+&\kappa\sin[x_{\mathrm{c}}^{(0)}+(s+3)\eta -u_{s+4}]  \notag \\
&-&\kappa\sin[x_{\mathrm{c}}^{(0)}+(s+1)\eta -u_{s+2}].  \label{v4}
\end{eqnarray}
To get $v_{s+4}$ to first order in $\kappa$, it is sufficient to write $%
u_{s+2}$ and $u_{s+4}$ in Eq. (\ref{v4}) to zero order in $\kappa$ by using
Eqs. (\ref{Mh4}) with $\kappa=0$: $u_{s+2}=-u_s$, $u_{s+4}=u_s$. Then, Eq. (%
\ref{v4}) becomes 
\begin{eqnarray}
v_{s+4}=v_s&-&\kappa\sin[u_s-x_{\mathrm{c}}^{(0)}-(s+3)\eta ]  \notag \\
&-&\kappa\sin[u_s+x_{\mathrm{c}}^{(0)}+(s+1)\eta ]  \label{v4m}
\end{eqnarray}
to first order in $\kappa$, as all expressions below. By iterating Eq. (\ref%
{v4m}), we get for $b>0$: 
\begin{eqnarray}
v_{4b+w}=v_w&-&\kappa\sum_{b^{\prime}=1}^b\{\sin[u_w-x_{\mathrm{c}%
}^{(0)}-(4b^{\prime}+w-1)\eta ]  \notag \\
&+&\sin[u_w+x_{\mathrm{c}}^{(0)}+(4b^{\prime}+w-3)\eta ]\}.  \label{v4bw}
\end{eqnarray}
Inserting Eq. (\ref{v4bw}) into Eq. (\ref{Fe}) and Taylor expanding around $%
v_w$ up to first order in $\kappa$, we obtain 
\begin{eqnarray}
F=&-&r^{-1}\sum_{w=0}^{3}e^{iw\pi /2}\sum_{b=0}^{l^{\prime}-1}\sin[x_{%
\mathrm{c}}^{(0)}+(4b+w)\eta-v_w]  \notag \\
&-&\kappa r^{-1}\sum_{w=0}^{3}e^{iw\pi /2}\sum_{b=1}^{l^{\prime}-1}\cos[x_{%
\mathrm{c}}^{(0)}+(4b+w)\eta-v_w]  \notag \\
&\times&\sum_{b^{\prime}=1}^b\{\sin[u_w-x_{\mathrm{c}}^{(0)}-(4b^{%
\prime}+w-1)\eta ]  \notag \\
&+&\sin[u_w+x_{\mathrm{c}}^{(0)}+(4b^{\prime}+w-3)\eta ]\} .  \label{Fe1}
\end{eqnarray}

In the first line of Eq. (\ref{Fe1}), the sum over $b$ vanishes identically
like a geometric sum since $4\eta =2\pi n^{\prime}k/l^{\prime}$, where, by
definition, $n^{\prime}/l^{\prime}=4/l$ with $(n^{\prime},l^{\prime})$
coprime integers. Thus, only the last three lines of Eq. (\ref{Fe1}) remain,
so that $F$ is of first order in $\kappa$, as expected for $l^{\prime}>1$
(SWC). Then, after some trigonometry, Eq. (\ref{Fe1}) can be written as
follows: 
\begin{equation}  \label{Few}
F=-\frac{\kappa}{r}\sum_{w=0}^{3}e^{iw\pi /2}\sin(u_w-\eta )(G_w+Q_w),
\end{equation}
where 
\begin{equation}  \label{Gw}
G_w=\sum_{b=1}^{l^{\prime}-1}\sum_{b^{\prime}=1}^b \cos[v_w+(4b^{%
\prime}-4b-2)\eta ],
\end{equation}
\begin{equation}  \label{Qw}
Q_w=\sum_{b=1}^{l^{\prime}-1}\sum_{b^{\prime}=1}^b \cos[2x_{\mathrm{c}%
}^{(0)}+2(w+2b+2b^{\prime}-1)\eta -v_w].
\end{equation}
A simple expression for $G_w$ in Eq. (\ref{Gw}) can be derived, 
\begin{equation}  \label{Gwe}
G_w=\frac{e^{i(v_w-2\eta )}}{2}\sum_{b=1}^{l^{\prime}-1}\sum_{b^{%
\prime}=1}^b e^{4i(b^{\prime}-b)\eta } +c.c.=\frac{l^{\prime}\sin(v_w)}{%
2\sin(2\eta )},
\end{equation}
after a lengthy but straightforward calculation of the geometric sums.

Similarly, $Q_w$ in Eq. (\ref{Qw}) can be explicitly calculated: 
\begin{eqnarray}
Q_w &=&
A_w\sum_{b=1}^{l^{\prime}-1}\sum_{b^{\prime}=1}^be^{4i(b^{\prime}+b)\eta
}+c.c.  \notag \\
&=& B_w\left( \frac{e^{8il^{\prime}\eta}-e^{8i\eta}}{e^{8i\eta}-1}-\frac{%
e^{4il^{\prime}\eta}-e^{4i\eta}}{e^{4i\eta}-1}\right) + c.c.,  \label{Qwe}
\end{eqnarray}
where 
\begin{equation}  \label{AB}
A_w=\frac{e^{i[2x_{\mathrm{c}}^{(0)}+2(w-1)\eta -v_w]}}{2},\ \ \ B_w=\frac{%
e^{4i\eta}A_w}{e^{4i\eta}-1}.
\end{equation}
Let us show that the expression (\ref{Qwe}) vanishes for $l^{\prime}>2$.
Since $4\eta =2\pi n^{\prime}k/l^{\prime}$, where $(n^{\prime},l^{\prime})$
are coprime integers (see also above), one always has $\exp
(4il^{\prime}\eta )=\exp (8il^{\prime}\eta )=1$. However, for $l^{\prime}>2$%
, $\exp (4i\eta )\neq 1$ and also $\exp (8i\eta )\neq 1$. Then, the
expression (\ref{Qwe}) vanishes.

For $l^{\prime}=2$, the integer $n^{\prime}k$ above is necessarily odd, so
that $\exp (4i\eta )=-1$ and $\exp (8i\eta )=1$. Using this in Eq. (\ref{Qwe}%
) with Eqs. (\ref{AB}), we get: 
\begin{equation}  \label{Qwe2}
Q_w=\cos [2x_{\mathrm{c}}^{(0)}+2(w-1)\eta -v_w].
\end{equation}

The quantity (\ref{Few}) for $l^{\prime}>2$ can now be written in a closed
form by using $r=4l^{\prime}$, Eq. (\ref{Gwe}), $Q_w=0$, and the expressions
of $(u_w,v_w)$ to zero order in $\kappa$, i.e., $u_1=-u_3=v_0$, $u_2=-u_0$, $%
v_1=-v_3=-u_0$, and $v_2=-v_0$. We then obtain Eq. (\ref{Feg}).

Similarly, using Eqs. (\ref{Gwe}) and (\ref{Qwe2}) in Eq. (\ref{Few}) for $%
l^{\prime}=2$ and $\eta =\pi /4$, we find that 
\begin{eqnarray}
F&=&\frac{\sqrt{2}\kappa}{8}[\cos(u_0)\sin(v_0)-i\sin(u_0)\cos(v_0)  \notag
\\
&+&\exp (2ix_{\mathrm{c}}^{(0)})\sin(u_0)\sin(v_0)  \notag \\
&+&i\exp (-2ix_{\mathrm{c}}^{(0)})\cos(u_0)\cos(v_0)].  \label{Feg2}
\end{eqnarray}

\begin{center}
\textbf{APPENDIX C}
\end{center}

We show here that a stochastic web is associated with the level set $H_{%
\mathrm{e}}=0$ for all effective Hamiltonians (\ref{He}), (\ref{Hea}), and (%
\ref{Heb}).

A stochastic web for the map (\ref{cMh4}) is a translationally invariant
chaotic region in phase space emerging from heteroclinic intersections of
the stable and unstable manifolds of neighboring hyperbolic fixed points of $%
M_{\gamma ,\eta ,r}$. These fixed points form a lattice in phase space which
is invariant under a rotation by angle $\gamma =\pi /2$ around $z=0$. In the
limit of $\kappa\rightarrow 0$, $M_{\gamma ,\eta ,r}$ is described by the
integrable effective Hamiltonian $H_{\mathrm{e}}$ and the stochastic web
reduces to the web skeleton, i.e., the union of straight lines connecting
the fixed points on the lattice. Because of the invariance of this lattice
under a rotation by $\gamma =\pi /2$, these lines form two perpendicular
sets, each set consisting of parallel lines in the direction of either the
stable ($\mathbf{U}_{-}$) or unstable ($\mathbf{U}_{+}$) eigenvector of the
linear-stability matrix $DM_{\gamma ,\eta ,r}$ at a fixed point. Therefore,
these eigenvectors, which are the limit $\kappa\rightarrow 0$ of the stable
and unstable manifolds, must be \emph{orthogonal}. The corresponding
eigenvalues have the form $\lambda_{\pm}=\exp(\pm\sigma )$, where $\sigma >0$
is a local Lyapunov exponent.

The fixed points $(u_0,v_0)$ in the $\kappa\rightarrow 0$ limit are
determined from Hamilton equations (\ref{hes}) with $\dot{u}=\dot{v}=0$: 
\begin{equation}  \label{fpe}
\left.\frac{\partial H_{\mathrm{e}}}{\partial u}\right|_{u_0,v_0}=\left.%
\frac{\partial H_{\mathrm{e}}}{\partial v}\right|_{u_0,v_0}=0.
\end{equation}
Denoting $\mathbf{R}_0=(u_0,v_0)^{\mathrm{T}}$, where T stands for
transpose, the linear stability of $\mathbf{R}_0$ under small perturbations $%
\delta\mathbf{R}_0$ is determined by linearizing Eqs. (\ref{hes}) around $%
\mathbf{R}_0$: 
\begin{equation}  \label{lhe}
\dot{\delta\mathbf{R}}_0=DH_e \delta\mathbf{R}_0,
\end{equation}
where $DH_e$ is the matrix 
\begin{equation}  \label{dhe}
DH_{\mathrm{e}}= 
\begin{pmatrix}
\frac{\partial ^{2}H_{\mathrm{e}}}{\partial u_0\partial v_0} & \frac{%
\partial ^{2}H_{\mathrm{e}}}{ \partial v_0^{2}} \\ 
&  \\ 
-\frac{\partial ^{2}H_{\mathrm{e}}}{\partial u_0^{2}} & \frac{\partial
^{2}H_{\mathrm{e}}}{ \partial u_0\partial v_0}%
\end{pmatrix}
.
\end{equation}
Assuming the time dependence $\delta\mathbf{R}_0(t)=\mathbf{U}\exp(\xi t)$
in Eq. (\ref{lhe}), we get the eigenvalue equation 
\begin{equation}  \label{ee}
DH_e \mathbf{U}=\xi \mathbf{U}.
\end{equation}
Now, as mentioned above, the eigenvectors $\mathbf{U}_{\pm}$, associated
with hyperbolic fixed points on a stochastic web, must be orthogonal and the
corresponding eigenvalues $\xi =\pm\sigma$ are real. Therefore, the real
matrix (\ref{dhe}) must be symmetric: 
\begin{equation}  \label{sc}
\frac{\partial ^{2}H_{\mathrm{e}}}{\partial v_0^{2}} = -\frac{\partial
^{2}H_{\mathrm{e}}}{\partial u_0^{2}}.
\end{equation}
In what follows, we show that the symmetry condition (\ref{sc}) is
equivalent to $H_{\mathrm{e}}=0$ in all the cases of Eqs. (\ref{He}), (\ref%
{Hea}), and (\ref{Heb}).

In the case of Eq. (\ref{He}), condition (\ref{sc}) is simply $-H_{\mathrm{e}%
}=H_{\mathrm{e}}$, implying $H_{\mathrm{e}}=0$.

In the case of Eq. (\ref{Hea}), let us first calculate from Eqs. (\ref{fpe})
the fixed points. Assuming $x_{\mathrm{c}}^{(0)}\neq 0,\pi /2$, these are
given by $u_0=\pi /2+a_1\pi$ and $v_0=\pi /2+a_2\pi$, for all integers $%
(a_1,a_2)$. Then, condition (\ref{sc}) reads: 
\begin{equation}  \label{sc2}
(-1)^{a_1}\cos(x_{\mathrm{c}}^{(0)})=(-1)^{a_2}\sin(x_{\mathrm{c}}^{(0)}).
\end{equation}
Using Eq. (\ref{sc2}) and the values above of $(u_0,v_0)$ in Eq. (\ref{Hea}%
), we find that $H_{\mathrm{e}}=0$. We note that Eq. (\ref{sc2}) also
implies that $|\tan (x_{\mathrm{c}}^{(0)})|=1$, i.e., the fixed points are
hyperbolic and form a web skeleton only if $x_{\mathrm{c}}^{(0)}=\pi /4$ and 
$a_1+a_2$ is even or $x_{\mathrm{c}}^{(0)}=3\pi /4$ and $a_1+a_2$ is odd.

Finally, in the case of Eq. (\ref{Heb}), condition (\ref{sc}) reads again $%
-H_{\mathrm{e}}=H_{\mathrm{e}}$, implying $H_{\mathrm{e}}=0$.


\begin{thebibliography}{99}
\bibitem{chaos} M.V. Berry, \emph{Regular and Irregular Motion} in Topics in
Nonlinear Mechanics, edited by S Jorna, Am. Inst. Ph. Conf. Proc. \textbf{46}%
, 16 (1978), and references therein; A.J. Lichtenberg and M.A. Lieberman, 
\emph{Regular and Chaotic Dynamics} (Springer-Verlag, New York, 1992), and
references therein.

\bibitem{bc} B.V. Chirikov, Phys. Rep. \textbf{52}, 263 (1979), and
references therein.

\bibitem{k} A.N. Kolmogorov, Dokl. Akad. Nauk USSR \textbf{98}, 572 (1954).

\bibitem{a} V.I. Arnol'd, Usp. Phys. Nauk \textbf{18}, 91 (1963) [Russ.
Math. Surv. \textbf{18}, 85 (1963)].

\bibitem{m} J. Moser, Nachr. Akad. Wiss. Gottingen Math. Phys. Kl. 2 \textbf{%
1}, 1 (1962).

\bibitem{jmg} J.M. Greene, J. Math. Phys. \textbf{20}, 1183 (1979).

\bibitem{mmp} R.S. MacKay, J.D. Meiss, and I.C. Percival, Physica D \textbf{%
13}, 55 (1984).

\bibitem{df} I. Dana and S. Fishman, Physica D \textbf{17}, 63 (1985).

\bibitem{dr} I. Dana and W.P. Reinhardt, Physica D \textbf{28}, 115 (1987).

\bibitem{a1} V. Arnold, Dokl. Akad. Nauk SSSR \textbf{156}, 9 (1964) [Soviet
Math. Dokl. \textbf{5}, 581 (1964)].

\bibitem{akn} V.I. Arnold, V.V. Kozlov, and A.I. Neishtadt, \emph{%
Mathematical Aspects of Classical and Celestial Mechanics} (Springer-Verlag,
Berlin, 1997).

\bibitem{kl} V. Kaloshin and M. Levi, SIAM Review \textbf{50}, 702 (2008),
and references therein.

\bibitem{wm} G.M. Zaslavskii, M. Yu. Zakharov, R. Z. Sagdeev, D. A. Usikov,
and A. A. Chernikov, Zh. Eksp. Teor. Fiz. \textbf{91}, 500 (1986) [Sov.
Phys. JETP \textbf{64}, 294 (1986)].

\bibitem{lw} A.J. Lichtenberg and B.P. Wood, Phys. Rev. A \textbf{39}, 2153
(1989).

\bibitem{y} Z. Yang, Phys. Rev. E \textbf{50}, 3582 (1994).

\bibitem{da} I. Dana and M. Amit, Phys. Rev. E \textbf{51}, R2731 (1995).

\bibitem{dh} I. Dana and T. Horesh, Lect. Notes Phys. \textbf{511}, 51
(1998).

\bibitem{d} I. Dana, Phys. Rev. Lett. \textbf{73}, 1609 (1994); see, in
particular, Fig. 1 in this paper.

\bibitem{prk} S. Pekarsky and V. Rom-Kedar, Phys. Lett. A \textbf{225}, 274
(1997).

\bibitem{dd} I. Dana and D.L. Dorofeev, Phys. Rev. E \textbf{72}, 046205
(2005).

\bibitem{jl} M.H. Johnson and B.A. Lippmann, Phys. Rev. \textbf{76}, 828
(1949).

\bibitem{note1} We see that the map (\ref{cMhb}) for $f(x)=-\sin(x)$ is
invariant under the simultaneous transformations $\gamma\rightarrow
\gamma+\pi$ and $\eta\rightarrow \eta+\pi$. Also, the map (\ref{Mh}) for $%
f(x)=-\sin(x)$ and $\gamma=\pi/2$ is invariant under the simultaneous
transformations $\eta\rightarrow \eta+\pi$ and $u_0\rightarrow u_0+\pi$.
Thus, by replacing $\exp(ij\pi/2)$ in Eq. (\ref{cMh4}) by $\exp(3ij\pi/2)$,
one just gets the map (\ref{cMh4}) with $u_0$ shifted by $\pi$.

\bibitem{note2} By definition of $l^{\prime}$, $4/l=n^{\prime}/l^{\prime}$
where $(n^{\prime},l^{\prime})$ are coprime integers. Thus, $l^{\prime}=1$
means that either $l=1$ or $l=4$. Since $\eta=2\pi k/l$, the case of $l=1$
for the map (\ref{cMh4}) is equivalent to that of $\eta=0$, previously
studied, and will therefore not be considered. For $l=4$, $\eta=\pi/2$ or $%
\eta=3\pi/2$. However, the case of $\eta=3\pi/2$ is equivalent to that of $%
\eta=\pi/2$ with $u_0$ shifted by $\pi$ (see note \cite{note1}).

\bibitem{sa1} S.S. Abdullaev, J. Phys. A \textbf{32}, 2745 (1999); J. Phys.
A \textbf{35}, 2811 (2002); Lect. Notes Phys. \textbf{691}, 1 (2006).

\bibitem{sa2} S.S. Abdullaev, Phys. Rev. E \textbf{76}, 026216 (2007).

\bibitem{dmcw} G.J. Duffy, A.S. Mellish, K.J. Challis, and A.C. Wilson,
Phys. Rev. A \textbf{70}, 041602(R) (2004).
\end{thebibliography}
\end{document}